\documentclass[12pt]{article}
\usepackage[utf8]{inputenc}

\usepackage{makeidx}         
\usepackage{graphicx}        
\usepackage{multicol}        
\usepackage{multirow}
\usepackage[bottom]{footmisc}
\usepackage{geometry} 
\usepackage{amsmath, amssymb, amsthm}
\usepackage{psfrag,epsf}
\usepackage{enumerate}
\usepackage{url} 
\usepackage{hyperref}
\usepackage{booktabs}
\usepackage{xcolor}
\usepackage{natbib}
\usepackage[italian,english]{babel}
\usepackage{authblk}
\usepackage{changes}
\usepackage{caption}
\usepackage{subcaption}
\usepackage{float}
\usepackage{placeins}
\usepackage{ulem}
\usepackage{cancel}

\hypersetup{
			colorlinks=true,
			linkcolor=blue,
            linktoc=page,
			anchorcolor=black,
			citecolor=blue,
			urlcolor=blue,
}

\def \bs{\mathbf}

\def\0{\mbox{\bf{0}}}
\def\bs{\mathbf{s}}\def\be{\mathbf{e}}

\usepackage[toc,page]{appendix}

\def \be{\begin{equation}}
\def \ee{\end{equation}}
\def \ber{\begin{eqnarray}}
\def \eer{\end{eqnarray}}
\def \berr{\begin{eqnarray*}}
\def \eerr{\end{eqnarray*}}
\def \bqmatrix{\begin{bmatrix}}
\def \eqmatrix{\end{bmatrix}}

\def \be{\begin{equation}}
\def \ee{\end{equation}}
\def \ber{\begin{eqnarray}}
\def \eer{\end{eqnarray}}
\def \berr{\begin{eqnarray*}}
\def \eerr{\end{eqnarray*}}
\def \bamatrix{\begin{pmatrix}}
\def \eamatrix{\end{pmatrix}}
\def \bqmatrix{\begin{bmatrix}}
\def \eqmatrix{\end{bmatrix}}

\def \argmin{{\rm arg\,min }}

\def \bs{\boldsymbol}

\def \qmo{``}

\def \qmcsp{'' }

\begin{document}
\title{\bf }
\author[1]{Beatrice Foroni\thanks{Corresponding author: beatrice.foroni@uniroma1.it}}
\affil[1]{MEMOTEF Department, Sapienza University of Rome}
\author[2]{Luca Merlo}
\affil[2]{Department of Human Sciences, European University of Rome}
\author[1]{Lea Petrella}

\title{Quantile and expectile copula-based hidden Markov regression models for the analysis of the cryptocurrency market}
\maketitle

\begin{abstract}
The role of cryptocurrencies within the financial systems has been expanding rapidly in recent years among investors and institutions. It is therefore crucial to investigate the phenomena and develop statistical methods able to capture their interrelationships, the links with other global systems, and, at the same time, the serial heterogeneity. 
 For these reasons, this paper introduces hidden Markov regression models for jointly estimating quantiles and expectiles of cryptocurrency returns using regime-switching copulas.
The proposed approach allows us to focus on extreme returns and describe their temporal evolution by introducing time-dependent coefficients evolving according to a latent Markov chain. Moreover to model their time-varying dependence structure, we consider elliptical copula functions defined by state-specific parameters. Maximum likelihood estimates are obtained via an Expectation-Maximization algorithm. The empirical analysis investigates the relationship between daily returns of five cryptocurrencies and major world market indices. 
\end{abstract}


\newpage

\section{Introduction}
Since the creation of Bitcoin more than a decade ago \citep{nakamoto2008bitcoin}, exchange volumes have expanded tremendously, facilitated by the speed of transaction and the lack of any central authority or financial intermediary. This exploitation has piqued the interest of policymakers, risk managers and academics in the peculiar characteristics of cryptocurrencies, which can serve both as an efficient payment method and an investment asset. The rapid development of many alternative crypto assets attempting to replicate the path of the Bitcoin has resulted in massive speculative manoeuvres, which have increased the willingness of investors and practitioners to comprehend this challenging financial market \citep{borri2019conditional}. Speculative periodic crypto-bubbles have repeatedly jeopardized the stability of financial markets and have been extensively studied, especially after the 2017 boom and the subsequent catastrophic crash in 2018 \citep{cheah2015speculative, cheung2015crypto, corbet2018exploring, agosto2020financial, xiong2020new}. 
At the beginning of 2020, the global COVID-19 pandemic and the prospect of a global recession exacerbated the volatility of cryptoassets, prompting practitioners to speculate about the impact of the pandemic on digital currency returns and volatility, as well as their effect on international stock markets. 
 Different aspects of cryptocurrencies, such as long-memory and efficiency \citep{duan2021dynamic, lopez2021efficiency, assaf2022multivariate}, hedging properties \citep{demir2020relationship, das2020does}, and relationships with other asset classes, have been investigated in the literature. In particular, empirical studies focused on whether cryptocurrencies could be used as optimal instruments to diversify investors' portfolios. \cite{conlon2020safe} analyzed if Bitcoin could be used as a safe-haven for the Standard $\&$ Poor's 500 (S$\&$P500). \cite{mariana2021bitcoin} similarly tested Bitcoin and Ethereum safe-haven properties, while \cite{corbet2020contagion} studied the potential increases in volatility or correlation between Bitcoin and traditional markets and commodities, such as gold and oil prices.
 In the last years, the financial literature on the empirical characteristics of digital currencies have documented the existence of stylized facts, such as volatility clustering, asymmetry and leptokurticity, and the presence of spillover effects within the crypto market and toward other global financial markets. 
 Consequently, it is indeed of utmost importance to be able to develop adequate statistical tools to take into account all these features. 
In the financial literature, Hidden Markov Models (HMMs, see \citealt{macdonald1997hidden, zucchini2016hidden}) have been successfully employed in understanding whether, and how, time series temporal evolution can be influenced by hidden variables during tranquil and crisis periods. Numerous HMMs applications can be found in asset allocation and stock returns, as discussed for example in \cite{mergner2008time, de2013dynamic, nystrup2017long, maruotti2019hidden}. In the context of cryptocurrencies time series, \cite{giudici2020hidden, caferra2021raised, pennoni2022exploring} and \cite{cremaschini2022stylized} consider latent Markov processes to analyze volatility clustering and serial heterogeneity of crypto returns. 
 Besides the clustering behavior of returns, the analysis of the dynamics of extreme returns is of the utmost importance for regulators and policymakers for modeling the entire distribution of returns while accounting for the well-known stylized facts, i.e., high kurtosis, skewness and serial correlation. In order to address these features, since the seminal work of \cite{koenker1978regression}, quantile regression represented a valid approach and a widely used technique in many empirical applications; see \cite{koenker2005quantile} and \cite{koenker2017handbook}. It allows to model the conditional quantiles of a response variable with respect to a set of covariates, providing a much more complete picture of the conditional distribution compared with traditional mean regression. Quantile models have been extensively applied in finance and economics for estimating Value at Risk (VaR) and quantile-based risk measures \citep{engle2004CAViaR, white2015var, bernardi2015bayesian, taylor2019forecasting, merlo2021, bottone2021unified, candila2023mixed}. 
 One of the most relevant extension related to quantile regression is provided by the expectile regression \citep{newey1987asymmetric}, which is a \qmo quantile-like\qmcsp generalization of the mean regression using an asymmetric squared loss function.
 Similarly to the former, the latter allows to represent the entire conditional distribution of a response variable and it possesses several advantages theoretically and computationally \citep{tzavidis2016longitudinal, alfo2017finite, nigri2022relationship}. Even though many quantile regression methods are now well consolidated in the literature, 
 few studies have been conducted on tail events and their link to traditional assets within the context of digital currencies. 
 \cite{borri2019conditional} employs the Conditional VaR to estimate the conditional tail risk in the cryptocurrency markets, indicating that cryptocurrencies are highly exposed to tail risk within the cryptocurrency markets but are disconnected from other global assets.
 \cite{ciner2022determinants} analyze cryptocurrency returns during highly volatile period of the COVID-19 pandemic using penalized quantile regressions, finding an impact of Gold and S$\&$P500 at the median of the crypto return distributions while, in the case of expectiles, \cite{foroni2023expectile} propose an expectile HMM for the analysis of Bitcoin daily returns.
 These proposals, however, are confined to the modeling of univariate financial time series only. When multiple cryptocurrencies returns are analyzed jointly, their dependence structure must be incorporated in the modeling framework in order to provide adequate risk control measures and produce effective asset allocation and diversification strategies. In these cases, taking into account the degree of association among different digital currencies, that cannot be detected by univariate methods, could be extremely important also for regulatory interventions. 
 To address all these features simultaneously, starting from the work of \cite{otting2021copula}, we develop hidden Markov regression models for jointly estimating conditional quantiles and expectiles of multiple cryptocurrencies returns, using regime dependent copulas in a multivariate framework. 
 Temporal evolution of returns is modeled through time-dependent coefficients in the regression model, evolving according to a discrete, homogeneous latent Markov chain and, at the same time, we consider elliptical copulas defined by state-specific parameters that take into account for the time-varying dependence structure of returns. 
 With this work, we unify in a common approach quantile and expectile HMMs with copulas, which allow us to pursue a two-fold goal. First, we incorporate within state-dependencies among the cryptocurrencies, which is crucial for investors whose investment portfolios contain a portion of crypto-assets as well as for policymakers whose role is to maintain the stability of financial markets. Second, we investigate the relationship among crypto and traditional financial assets at different volatility states, and in lower and upper tails of cryptocurrency return distributions. 
 The estimation is carried out in a Maximum Likelihood (ML) framework using, respectively, the Asymmetric Laplace \citep{yu2001bayesian} and Asymmetric Normal \citep{waldmann2017bayesian} distributions for quantiles and expectiles as working likelihoods through suitable Expectation-Maximization (EM) algorithms. 
 The good performances of our methods are illustrated through a simulation study generating observations from a bivariate two-state HMM under different sample sizes, error distributions and copula functions.
 The real data analysis considers daily returns from July 2017 until December 2022, which comprise numerous events that heavily impacted financial stability, as the crypto currency bubble crisis in 2017-2018, the COVID-19 outbreak in 2020, Biden's election at the USA presidency in November 2020 and the Russian invasion of Ukraine at the beginning of 2022. As for the choice of the dependent variables, we select five cryptocurrencies, namely Bitcoin (BTC), Ethereum (ETH), Ripple (XRP), Litecoin (LTC), and Bitcoin Cash (BCH), following the criteria adopted in \cite{pennoni2022exploring}. Digital currencies are modeled as functions of major stock and global market indices, including S$\&$P500, S$\&$P US Treasury Bond, US dollar index, WTI Crude Oil and Gold COMEX daily closing prices. Our results show that cryptocurrency returns exhibit a clear temporal clustering behavior in calm and turbulent periods, and the association with traditional financial assets is strong at extreme tails of returns distribution, especially with S$\&$P500, S$\&$P US Treasury Bond and Gold. 
 
The rest of the paper is organized as follows. Section \ref{sec:quant_exp} briefly reviews univariate quantile and expectile regressions. In Section \ref{sec:method} we describe the proposed models, the EM algorithms for estimating the model parameters and the computational aspects of the estimation procedure. In Section \ref{sec:sim}, we evaluate the performance of our proposal in a simulation study. 
 Section \ref{sec:emp} is devoted to the empirical analysis and discusses the results obtained while Section \ref{sec:conclusions} concludes. 

\section{Preliminaries on quantile and expectile regressions}\label{sec:quant_exp}
In order to better explain the proposed models, in this section we briefly revise the univariate quantile and expectile regressions.

 For a continuous response variable, quantile and expectile regressions provide a much more flexible approach and complete picture of the conditional distribution of the response than classical regression models targeting the mean. The former, introduced by \cite{koenker1978regression}, can be considered as a generalisation of median regression, while the latter, proposed by \cite{newey1987asymmetric}, can be thought as a generalization of mean regression based on asymmetric least-squares estimation. More generally, both quantiles and expectiles can be embedded in a common framework within the wider class of generalized quantiles defined as the minimizers of an asymmetric $l$-power loss function.  
 Formally, let $\tau \in (0,1)$ and consider the following asymmetric loss function
\be\label{eq:lf}
\omega_{l,\tau}(u) = |u|^l \cdot |\tau - \mathbb{I}(u < 0)|, \ \ u \in \mathbb{R} 
\ee
where $\mathbb{I}(\cdot)$ denotes the indicator function.

 Given a set of covariates $\bs X = \bs x$, it is easy to see that when $l=1$, the conditional quantile of order $\tau$ of a continuous response $Y$ is defined as 
\be\label{eq:q}
q_{\bs x}(\tau) = \underset{m \in \mathbb{R}}{\argmin} \, \mathbb{E}[\omega_{1,\tau}(Y - m_{\bs x}(\tau))]
\ee 
meanwhile, for $l=2$, the $\tau$-th conditional expectile of $Y$ is defined as
\be\label{eq:e}
e_{\bs x}(\tau) = \underset{m \in \mathbb{R}}{\argmin} \, \mathbb{E}[\omega_{2,\tau} (Y - m_{\bs x}(\tau))].
\ee
In particular, $q_{\bs x}(\tau)$ and $e_{\bs x}(\tau)$ with $\tau = \frac{1}{2}$ correspond respectively to the conditional median and the mean of $Y$ given covariates $\bs x$. When $\tau \neq \frac{1}{2}$, both methods allow to target the entire conditional distribution of the response. In practice, quantiles have a more intuitive interpretation than expectiles even if they target essentially the same part of the distribution of interest. However, despite the popularity and the easy interpretability of the former, the latter offer some advantage: (a) we gain uniqueness of the ML solutions which is not granted in the quantile context; (b) from a computational standpoint, since the squared loss function $\omega_{2,\tau}(\cdot)$ is differentiable, $e_{\bs x}(\tau)$ can be estimated by efficient Iterative Reweighted Least Squares (IRLS), in contrast to algorithms used for fitting quantile regression models.

 From a likelihood perspective, both methods have been implemented in a ML approach by exploiting the relationship between the minimization of the loss function in \eqref{eq:lf} and the maximization of a likelihood function formed by combining independently distributed densities with kernel function $\omega_{l,\tau}(\cdot)$. That is: 
\be\label{eq:dens}
f(y; \mu, \sigma, \tau) = B_{l,\tau} (\sigma) \exp \left[-\omega_{l,\tau}\left( \frac{y - \mu}{\sigma} \right) \right]
\ee
where $\mu \in \mathbb{R}$ is a location parameter, $\sigma > 0$ is a scale parameter and $B_{l,\tau} (\sigma)$ is a normalizing constant that ensures the density integrates to one.

In the case of quantiles i.e. for $l=1$, the density in \eqref{eq:dens} reduces to the Asymmetric Laplace (AL) distribution introduced by \cite{yu2001bayesian}, $f_{AL}(y; \mu, \delta, \tau)$, where $\mu$ coincides with the $\tau$-th quantile of $Y$ with $B_{1,\tau} (\sigma) = \frac{\tau (1-\tau)}{\sigma}$. As regards to expectiles i.e. when $l=2$, \eqref{eq:dens} corresponds to the Asymmetric Normal (AN) distribution proposed by \cite{gerlach2015bayesian} and \cite{waldmann2017bayesian}, $f_{AN}(y; \mu, \sigma, \tau)$, and $\mu$ is the $\tau$-th expectile of $Y$ with $B_{2,\tau} (\sigma) = \frac{2\sqrt{\tau (1-\tau)}}{\sqrt{\pi \sigma^2} (\sqrt{\tau} + \sqrt{1-\tau})}$. It is easy to verify that in both cases, as discussed in \cite{yu2001bayesian} and \cite{waldmann2013bayesian} respectively, the minimization of the respective expected loss functions $\omega_{1,\tau}(\cdot)$ and $\omega_{2,\tau}(\cdot)$ is equivalent, in terms of parameter estimates, to the maximization of the likelihood functions associated with the AL and AN densities.
 In the following section, we extend quantile and expectile regressions to the HMM setting for the analysis of multivariate time series by considering elliptical copulas.

\section{Methodology}\label{sec:method}
In this section we formally introduce the quantile and expectile copula-based hidden Markov regression models. We then build suitable EM algorithms for ML estimation using the AL and AN distributions as working likelihoods for the proposed models. 
 Formally, let $\{ S_t \}_{t=1}^T$ be a latent, homogeneous, first-order Markov chain defined on the discrete state space $\{1,\dots,K\}$. Let $\pi_k = Pr(S_1=k)$ be the initial probability of state $k$, $k = 1,\dots,K$, and $\pi_{k|j} = Pr(S_{t+1}=k | S_t=j)$, with $\sum_{k=1}^K \pi_{k|j} = 1$ and $\pi_{k|j} \geq 0$, denote the transition probability between states $j$ and $k$, that is, the probability to visit state $k$ at time $t+1$ from state $j$ at time $t$, $j,k = 1,\dots,K$ and $t=1,\dots,T$. More concisely, we collect the initial and transition probabilities in the $K$-dimensional vector $\bs \pi$ and in the $K \times K$ matrix $\bs \Pi$, respectively.
 
Let $\bs Y_t = (Y_{t,1},\dots,Y_{t,d})$ denote a continuous $d$-dimensional dependent variable and $\bs x_t= (1, x_{t,2},\dots, x_{t,p})$ be a $p$-dimensional vector of fixed covariates, with the first element being the intercept, at time $t=1,\dots,T$. Finally, let $\bs \tau = (\tau_1, \dots, \tau_d)$ denote a $d$-dimensional vector of fixed scalars with $\tau_j \in (0,1)$, $j=1,\dots,d$. 
 As mentioned in the Introduction, our goal is to jointly model the univariate component-wise quantiles (expectiles) of the conditional distribution of the vector $\bs Y_t$ given $\bs x_t$ and $S_t = k$, capturing the possible dependence structure between cryptocurrencies returns'. To this end, we construct a multivariate state-dependent distribution allowing for within-state correlation among the elements in $\bs Y_t$ by using a copula-based approach. Denoting with $F_{Y_{t,j}} (y_{t,j} | \bs x_t, S_t = k)$, $j=1,\dots,d$, the distribution functions of the marginals, the state-dependent multivariate distribution of $\bs Y_t$ given covariates and $S_t = k$, is defined by 
\be 
F_{\bs Y_{t}} (\bs y_{t} | \bs x_t, S_t = k) = C(F_{Y_{t,1}} (y_{t,1} | \bs x_t, S_t = k), \dots, F_{Y_{t,d}} (y_{t,d} | \bs x_t, S_t = k); \bs \eta_k),
\ee
where $C(\cdot; \bs \eta_k)$ is a $d$-variate copula with time-varying parameter vector $\bs \eta_k$ that evolves over time according to the hidden process $S_t$ and takes one of
the values in the set $\{\bs \eta_1, \dots, \bs \eta_K \}$. As a consequence of Sklar's theorem (\citealt{sklar1959fonctions}), when the marginal distribution functions are continuous and strictly increasing, the joint density $f_{\bs Y_t}(\bs y_{t}| \bs{x}_t, S_t = k)$ can be written as
\be\label{eq:jdensc}
f_{\bs Y_t}(\bs y_{t}| \bs{x}_t, S_t = k) = \prod_{j=1}^d f_{Y_{t,j}}(y_{t,j} | \bs{x}_t, S_t = k) \cdot c(u_1, \dots, u_d; \bs \eta_k)
\ee
where $u_j = F_{Y_{t,j}} (y_{t,j} | \bs x_t, S_t = k)$ and $c(\cdot; \bs \eta_k)$ is the copula density
\be
c (\cdot; \bs \eta_k) = \frac{\partial^d C(\cdot; \bs \eta_k)}{\partial F_1 \cdots \partial F_d}.
\ee

Estimation of model parameters can be pursued using a ML approach. Specifically, to describe the conditional distribution of each response $Y_{t,j}$, $j = 1,\dots,d$, 
 we use the density in \eqref{eq:dens} whose probability density function is now given by
\be\label{eq:densc}
f_{Y_{t,j}}(y_{t,j} | \bs{x}_t, S_t = k) = B_{l,\tau_j} (\sigma_{j,k}) \exp \left[-\omega_{l,\tau_j}\left( \frac{y_{t,j} - \mu_{t,j,k}}{\sigma_{j,k}} \right) \right],
\ee
where the location parameter $\mu_{t,j,k}$ is defined by the following linear model:
\be\label{eq:linmod}
\mu_{t,j,k} (\tau_j) = \bs x_t \bs\beta_{j,k}(\tau_j), \quad j=1,\dots,d,
\ee
with $\bs \beta_{j,k} (\tau_j)$ being the $p$-dimensional state-specific regression parameters that assumes one of the values $\{\bs\beta_{j,1}(\tau_j), \dots, \bs\beta_{j,K}(\tau_j)\}$ depending on the outcome of the Markov chain $S_t$. As described in Section \ref{sec:quant_exp}, one of the advantages of using the distribution in \eqref{eq:densc} is that $\mu_{t,j,k}$ coincides with the $\tau_j$-th conditional quantile of $Y_{t,j}$ when $l=1$, which reduces to the AL distribution with $B_{l,\tau_j} (\sigma_{j,k}) = \frac{\tau_j (1-\tau_j)}{\sigma_{j,k}}$. Similarly, when $l=2$, $\mu_{t,j,k}$ represents the $\tau_j$-th conditional expectile of $Y_{t,j}$ and \eqref{eq:densc} corresponds to the AN distribution with $B_{l,\tau_j} (\sigma_{j,k}) = \frac{2\sqrt{\tau_j (1-\tau_j)}}{\sqrt{\pi \sigma_{j,k}^2} (\sqrt{\tau_j} + \sqrt{1-\tau_j})}$.
 These results hold true for all $k=1,\dots,K$ and $j=1,\dots,d$. In particular, \eqref{eq:jdensc}-\eqref{eq:linmod} define the proposed Copula Quantile Hidden Markov Model (CQHMM) when $l=1$ and the Copula Expectile Hidden Markov Model (CEHMM) when $l=2$. With respect to the current literature, the proposed method reduces, respectively, to $d$ separate linear quantile and expectile HMMs by \cite{farcomeni2012quantile} and \cite{foroni2023expectile} for $l=1$ and $l=2$ when assuming conditional independence between the responses, i.e., under the independence copula. To model the within state-dependence between the elements of $\bs Y_t$ we focus on the family of elliptical copulas which are derived from elliptically countered distributions; in particular we choose the Gaussian and the t distributions. Specifically, distribution functions of Gaussian and t copulas can be written as

\be\label{eq:copulag_cdf}
C^G(\bs u; \bs \Omega^{\Phi}) = \Phi_d(\Phi^{-1}(u_1),\dots,\Phi^{-1}(u_d); \bs \Omega^{\Phi})
\ee
 and
\be\label{eq:copulat_cdf}
C^t(\bs u; \bs \Omega^{\Psi}, \nu) = \Psi_d(\Psi^{-1}(u_1; \nu),\dots,\Psi^{-1}(u_d, \nu); \bs \Omega^{\Psi}, \nu),
\ee
where $\Phi_d$ and $\Psi_d$ denote the joint distribution functions for the $d$-variate normal and t distributions,  with correlation matrices $\bs \Omega^{\Phi}$ and $\bs \Omega^{\Psi}$ respectively. $\Phi^{-1}$ and $\Psi^{-1}$ are the inverse distribution functions of the univariate standard distributions, and as regards the t copula $\nu$ represents the degrees of freedom parameter, which we impose $\nu > 2$ to obtain a finite value for the variance. It is easy to see that as $\nu \rightarrow \infty$, the Gaussian copula in \eqref{eq:copulag_cdf} may be thought of as a limiting case of the t copula.
	
\subsection{Likelihood inference}\label{subsec:inference}
In this section we consider a ML approach to make inference on model parameters. As is common for HMMs, and for latent variable models in general, we develop EM algorithms (\citealt{baum1970maximization}) to estimate the parameters of the methods proposed. As we will show in the following, the EM algorithms for fitting the proposed CQHMM and CEHMM have a similar structure and thus, so we present a general framework to avoid redundancies. To ease the notation, unless specified otherwise, hereinafter we also omit the vector $\bs \tau$ representing the quantile (expectile) indices, yet all model parameters are allowed to depend on it.
\\
Let us denote with $\bs \theta = (\bs \beta_1, \dots, \bs \beta_K, \sigma_1, \dots, \sigma_K, \bs \pi, \bs \Pi, \bs \eta_1, \dots, \bs \eta_K)$ the vector of all model parameters. For the Gaussian copula, $\bs \eta_k$ comprises the elements of the state-specific correlation matrices $\bs \Omega^\Phi_k$, while for the t copula $\bs \eta_k = (\bs \Omega^\Psi_k, \nu_k)$, $k = 1,\dots,K$.

Thus, for a given number of hidden states $K$, the EM algorithm runs on the complete log-likelihood function of the models introduced, which is defined as	 
\be\label{eq:completel}
\ell_c(\bs \theta) = \sum_{k=1}^K {\gamma_1}(k)\log\pi_k + \sum_{t=1}^T\sum_{k=1}^K \sum_{j=1}^K {\xi_t}(j,k) \log \pi_{k|j} + \sum_{t=1}^T\sum_{k=1}^K {\gamma_t}(k)\log f_{\bs Y_t}(\bs y_t| \bs{x}_t, S_t = k),
\ee
where the joint density $f_{\bs Y_t}(\bs y_t| \bs{x}_t, S_t = k)$ is given in \eqref{eq:jdensc}, ${\gamma_t}(k)$ denotes a dummy variable equal to $1$ if the latent process is in state $k$ at occasion $t$ and 0 otherwise, and ${\xi_t}(j,k)$ is a dummy variable equal to $1$ if the process is in state $j$ in $t-1$ and in state $k$ at time $t$ and $0$ otherwise. 

To estimate $\bs \theta$, the algorithm iterates between two steps, the E- and M-steps, until convergence, as outlined below.

\subsubsection*{E-step:} 
In the E-step, at the generic $(h+1)$-th iteration, the unobservable indicator variables ${\gamma_t}(k)$ and ${\xi_t}(j,k)$ in \eqref{eq:completel} are replaced by their conditional expectations given the observed data and the current parameter estimates $\bs \theta^{(h)}$. To compute such quantities we require the calculation of the probability of being in state $k$ at time $t$ given the observed sequence 
\be\label{gamma}
\gamma_t^{(h)}(k) = P_{\bs \theta^{(h)}}(S_t = k |\bs y_1,\dots,\bs y_T ) = \frac{a_{t,k} b_{t,k}}{\sum_{k'=1}^K a_{t,k'} b_{t,k'}}
\ee
and the probability that at time $t-1$ the process is in state $j$ and then in state $k$ at time $t$, given the observed sequence
\be\label{xi} 
\xi_t^{(h)}(j,k) = P_{\bs \theta^{(h)}}(S_{t-1} = j, S_t = k |\bs y_1,\dots,\bs y_T ) = \frac{a_{t-1,k} \pi_{k|j} f (\cdot) b_{t,k}}{\sum_{k'=1}^K a_{T,k'}},
\ee
where $f (\cdot)$ is the AL density for the CQHMM or the AN distribution for the CEHMM in \eqref{eq:densc}, and $a_{t,k}$ and $b_{t,k}$ represent the forward and backward probabilities of each model. Such quantities 
can be efficiently obtained using the well-known Forward-Backward algorithm; see \cite{baum1970maximization} and \cite{welch2003hidden}. Then, we use these to calculate the conditional expectation of the complete log-likelihood function in \eqref{eq:completel} given the observed data and the current estimates:
\be\label{eq:estep}
\begin{aligned}
Q(\bs \theta|\bs \theta^{(h)}) =& \sum_{k=1}^K \gamma^{(h)}_1(k)\log\pi_k + \sum_{t=1}^T\sum_{k=1}^K \sum_{j=1}^K \xi^{(h)}_t(j,k) \log \pi_{k|j} + \\ &\sum_{t=1}^T\sum_{k=1}^K \gamma^{(h)}_t(k)\log f_{\bs Y_t}(\bs y_t| \bs{x}_t, S_t = k).
\end{aligned}
\ee

\subsubsection*{M-step:} 
In the M-step we maximize $Q(\bs \theta |\bs \theta^{(h)})$ in \eqref{eq:estep} with respect to $\bs \theta$ to obtain the update parameter estimates $\bs \theta^{(h+1)}$. 
 Formally, the initial probabilities $\pi_k$ and transition probabilities $\pi_{k|j}$ are updated using:
\be
\pi^{(h+1)}_k = \gamma^{(h)}_1(k), \quad k = 1,\dots,K
\ee
and
\be
\pi^{(h+1)}_{k|j} = \frac{\sum_{t=1}^T \xi^{(h)}_t(j,k)}{\sum_{t=1}^T \sum_{k=1}^K \xi^{(h)}_t(j,k)}, \quad j,k = 1,\dots,K.
\ee

To reduce the computational difficulty of the algorithm, we then update the state-dependent regression and copula parameters by adopting the Inference Functions for Margins (IFM) method proposed in \cite{joe1996estimation}. This entails a two-stage estimation procedure that first estimates the regression parameters of each component of $\bs Y_t$, $(\bs \beta_{j,k}, \sigma_{j,k})$, and it then obtains the parameters of the copula function, $\bs \eta_k$. To implement this step, we write the last term in \eqref{eq:estep} as follows:
\be\label{eq:regpar}
\begin{aligned}
& \sum_{t=1}^T\sum_{k=1}^K \gamma^{(h)}_t(k) \log f_{\bs Y_t}(\bs y_t| \bs{x}_t, S_t = k) = \\
& \sum_{t=1}^T\sum_{k=1}^K \gamma^{(h)}_t(k) \Big( \log c(F_{Y_{t,1}} (y_{t,1} | \bs x_t, S_t = k), \dots, F_{Y_{t,d}} (y_{t,d} | \bs x_t, S_t = k); \bs \eta_k) \Big) + \\
&\sum_{t=1}^T\sum_{k=1}^K \gamma^{(h)}_t(k) \Big( \sum_{j=1}^d \log f_{Y_{t,j}}(y_{t,j} | \bs{x}_t, S_t = k)  \Big),
\end{aligned}
\ee
where $f_{Y_{t,j}}(y_{t,j} | \bs{x}_t, S_t = k)$ and $F_{Y_{t,j}} (y_{t,j} | \bs x_t, S_t = k)$ are the univariate conditional density and distribution function for each component $j$ and each state $k$, respectively. In the CQHMM, these correspond to the density and distribution functions of the AL distribution, whereas, in the CEHMM, they represent the density and distribution functions of the AN distribution.

In the first step, we update the regression and scale parameters of each univariate conditional distribution by maximizing the second term in \eqref{eq:regpar}. Specifically, the estimate of the regression parameters are updated as follows

\be\label{eq:beta_quant}
\bs \beta^{(h+1)}_{j,k} = \underset{\bs \beta}{\argmin} \sum_{t=1}^T \gamma^{(h)}_t(k) \omega_{l,\tau_j}(y_{t,j} - \bs x'_t \bs \beta).
\ee
In the case of quantiles, a solution to \eqref{eq:beta_quant} can be obtain by fitting a linear quantile regression with weights $\gamma^{(h)}_t(k)$ for $l=1$. As for expectiles, due to the differentiability of the loss function $\omega_{2,\tau_j} (\cdot)$, $\bs \beta^{(h+1)}_{j,k}$ can be efficiently computed using IRLS for cross-sectional data with appropriate weights. Similarly, the scale parameters of each marginal distribution can be obtained using the following M-step update formulas for the CQHMM
\be
\sigma_{j,k}^{(h+1)} = \frac{1}{\sum_{t=1}^T \gamma^{(h)}_t(k)} \sum_{t=1}^T\sum_{k=1}^K \gamma^{(h)}_t(k) \omega_{1,\tau_j} (y_{t,j} - \bs x'_t \bs \beta^{(h+1)}_{k,j})
\ee
and for the CEHMM
\be
\sigma_{j,k}^{2(h+1)} = \frac{2}{\sum_{t=1}^T \gamma^{(h)}_t(k)} \sum_{t=1}^T \gamma^{(h)}_t(k) |\tau_j - \mathbb{I}(y_{t,j} < \bs x'_t \bs \beta^{(h+1)}_{j,k})| (y_{t,j} - \bs x'_t \bs \beta^{(h+1)}_{j,k})^2.
\ee

 In the second step, given the estimates of $\bs \beta^{(h+1)}_{j,k}$ and $\sigma_{j,k}^{(h+1)}$ for all components $j = 1,\dots,d$ and for all hidden states, we compute the copula parameters for $k = 1,\dots,K$ as follows:
\be\label{eq:copulamax}
\bs \eta^{(h+1)}_k = \underset{\bs \eta}{\argmin} \sum_{t=1}^T {\gamma^{(h)}_t}(k) \log c(F^{(h+1)}_{Y_{t,1}} (y_{t,1} | \bs x_t, S_t = k), \dots, F^{(h+1)}_{Y_{t,d}} (y_{t,d} | \bs x_t, S_t = k); {\bs \eta}),
\ee
where $F^{(h+1)}_{Y_{t,j}} (y_{t,j} | \bs x_t, S_t = k)$ denotes the univariate conditional distribution of the $j$-th response evaluated at $(\bs \beta^{(h+1)}_{j,k}, \sigma_{j,k}^{(h+1)})$.
 
 For the Gaussian copula, the update of the state-dependent correlation matrix $\bs \Omega^\Phi_k$ is obtained by maximizing \eqref{eq:copulamax}, which has a closed form expression. This coincides with the weighted sample covariance matrix of the pseudo-observations $u_j = F^{(h+1)}_{Y_{t,j}} (y_{t,j} | \bs x_t, S_t = k)$, $j=1,\dots,d$, with weights ${\gamma^{(h)}_t}(k)$. In the case of the t copula, in order to reduce the dimensionality of the optimization problem in \eqref{eq:copulamax}, we update the correlation matrix $\bs \Omega^\Psi_k$ for a fixed value of $\nu_k$. This is equivalent to estimating the scale matrix of a multivariate t distribution with zero mean vector, which can be done in closed form exploiting the hierarchical representation of the t distribution (\citealt{liu1997ml}). 
 Once we updated $\bs \Omega^\Psi_k$, the objective function in \eqref{eq:copulamax} is only a function of $\nu_k$ and, thus, the related computing is fast.\\

The E- and M- steps are alternated until convergence, that is when the observed likelihood between two consecutive iterations is smaller than a predetermined threshold. In this paper, we set this threshold criterion equal to $10^{-6}$. 
\\
Following \cite{maruotti2021hidden} and \cite{merlo2022quantile}, for fixed $\bs \tau$ and $K$ we initialize the EM algorithm by providing the initial states partition, $\{ S_t^{(0)} \}_{t=1}^T$, according to a Multinomial distribution with probabilities $1/K$. From the generated partition, the elements of $\bs \Pi^{(0)}$ are computed as proportions of transition, while we obtain $\bs \beta^{(0)}_{j,k}$ and $\sigma^{(0)}_{j,k}$ by fitting univariate mean and median regressions on the observations within state $k$ for the CQHMM and CEHMM, respectively. The state-dependent correlation matrices of the copula are set equal to the empirical correlation matrices computed on observations in the $k$-th state, while the initial value for the degrees of freedom of the t copula is $\nu^{(0)} = \nu^{(0)}_k = 5$ for all $k=1,\dots,K$. To deal with the possibility of multiple roots of the likelihood equation and better explore the parameter space, 
 we fit the proposed CQHMM and CEHMM using a multiple random starts strategy with different starting partitions and retain the solution corresponding to the maximum likelihood value. 

Once we computed the ML estimates of the model parameters, to estimate the standard errors we employ the parametric bootstrap scheme of \cite{visser2000confidence}. 
\section{Simulation Studies}\label{sec:sim} 
In this section we evaluate the performance of the proposed CQHMM and CEHMM via simulated data under different scenarios. In particular, we assess the ability of recovering the true regression parameter values and the clustering accuracy.
We consider a two-dimensional response variable ($d=2$), two sample sizes ($T = 500, \ T = 1000$) and one explanatory variable $X_t \sim \mathcal{N}(0,1)$. Observations are drawn from a bivariate two-states HMM using the following data generating process for $t = 1,\dots,T$,
\be\label{eq:gen_proc}
\bs Y_t = \bs X_t \bs \beta_k + \bs \epsilon_{t,k}, \quad S_t = k
\ee
where $\bs X_t = (1, X_t)'$ and the true values of the regression parameters are
$$\bs \beta_1 = \begin{pmatrix}
	-2 & 3 \\
	1 & -2
\end{pmatrix} \quad  \textnormal{and} \quad  \bs \beta_2 =  \begin{pmatrix}
	3 & -2 \\
	-2 & 1
\end{pmatrix}.$$
We consider three distributions for the error term in (\ref{eq:gen_proc}). In the first case, we generate $\bs \epsilon_{t,k}$ from a multivariate Gaussian distribution with zero mean vector and variance-covariance matrix equal to $\bs \Omega_{k}$. In the second one, $\bs \epsilon_{t,k}$ is generated from a multivariate t distribution with $5$ degrees of freedom, zero mean vector and scale matrix $\bs \Omega_{k}$. In the third scenario, $\bs \epsilon_{t,k}$ follows a multivariate skew-\textit{t} distribution with $5$ degrees of freedom, skewness parameters $\bs \alpha = (-2,2)$ and scale matrix $\bs \Omega_{k}$. The state-specific covariance matrices for the two states are equal to:
$$
\bs \Omega_{1} = \begin{pmatrix}
	1 & 0.2 \\
	0.2 & 1
\end{pmatrix}
 \quad \textnormal{and} \quad 
 \bs \Omega_{2} = \begin{pmatrix}
	1 & 0.7 \\
	0.7 & 1
\end{pmatrix}.$$
Finally, the matrix of transition probabilities is $\bs \Pi = \begin{pmatrix}
	0.9 & 0.1 \\
	0.1 & 0.9
\end{pmatrix}.$

 We fit the proposed CQHMM and CEHMM at three $\tau$ levels, i.e., $\tau = \{0.1, 0.5, 0.9 \}$ with, $\tau = \tau_j$, $j=1,2$, by assuming the Gaussian and t copulas for each scenario considered. In order to assess the validity of the models, we compute the absolute bias and standard errors associated to the state-specific regression coefficients $\bs \beta_1$ and $\bs \beta_2$, averaged over 100 Monte Carlo replications, for each combination of sample size, error distribution and copula function.

 In Tables \ref{tab:sim_ndist_q}, \ref{tab:sim_tdist_q} and \ref{tab:sim_stdist_q} we report the estimation results for the CQHMM regression parameters for each combination of copulas and error distributions. In this case we observe that the precision of the estimates is higher at the center of the distribution rather than on the tails, especially for the fat-tailed t and skew t distributions, mainly due to the reduced number of observations at extreme quantile levels. Some improvements can be observed when the parameters of the skew t case are estimated with the copula t rather than the Gaussian. Moving onto the estimation results of the CEHMM (Tables \ref{tab:sim_ndist_e}, \ref{tab:sim_tdist_e}, \ref{tab:sim_stdist_e}), we observe that the recovery performance of the data generation parameters is always very satisfactory, no matter the sample size, the copula or the generating process considered. Even when we generate from a skew-\textit{t} distribution we obtain good results on extreme expectiles, despite having a reduced amount of observations at the tails of the distribution. When we increase the sample size ($T=1000$) we observe that standard deviation tends to decrease, as expected. Overall, the expectile model achieves lower bias and standard errors with respect to the quantile model, especially at the tails of non-Gaussian distributions.
 
\begin{figure}[htbp]
\centering
\begin{subfigure}[b]{1.0\textwidth}
\includegraphics[width=.5\linewidth]{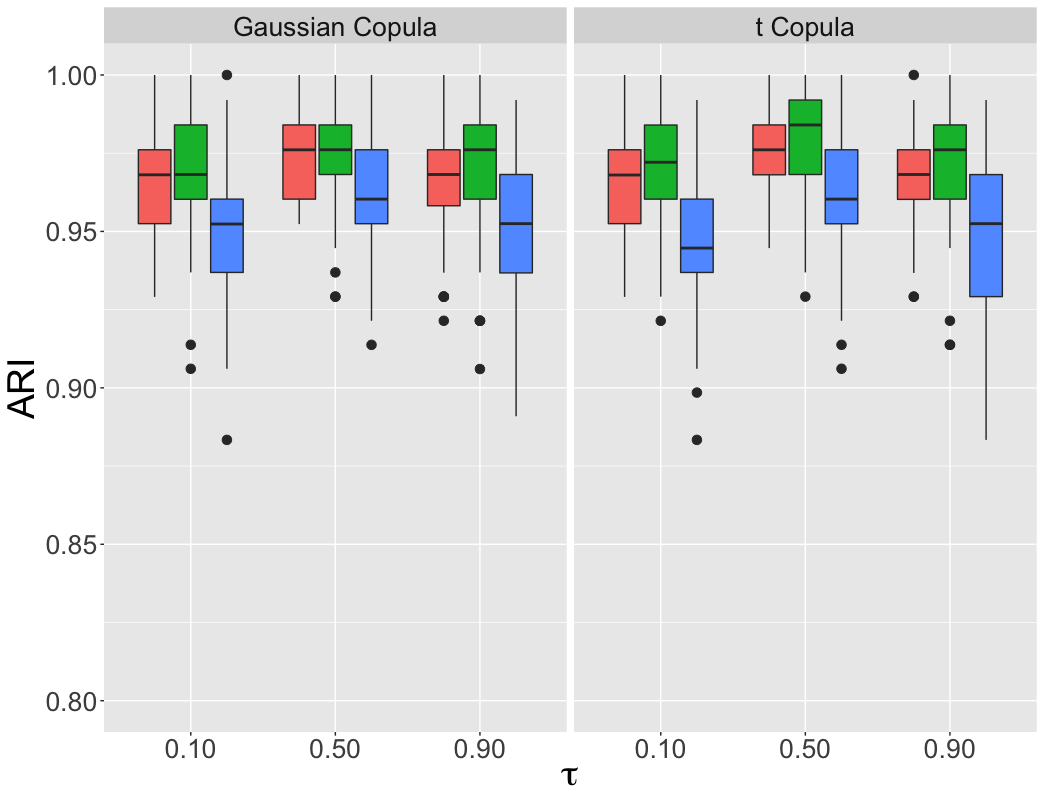}
\includegraphics[width=.5\linewidth]{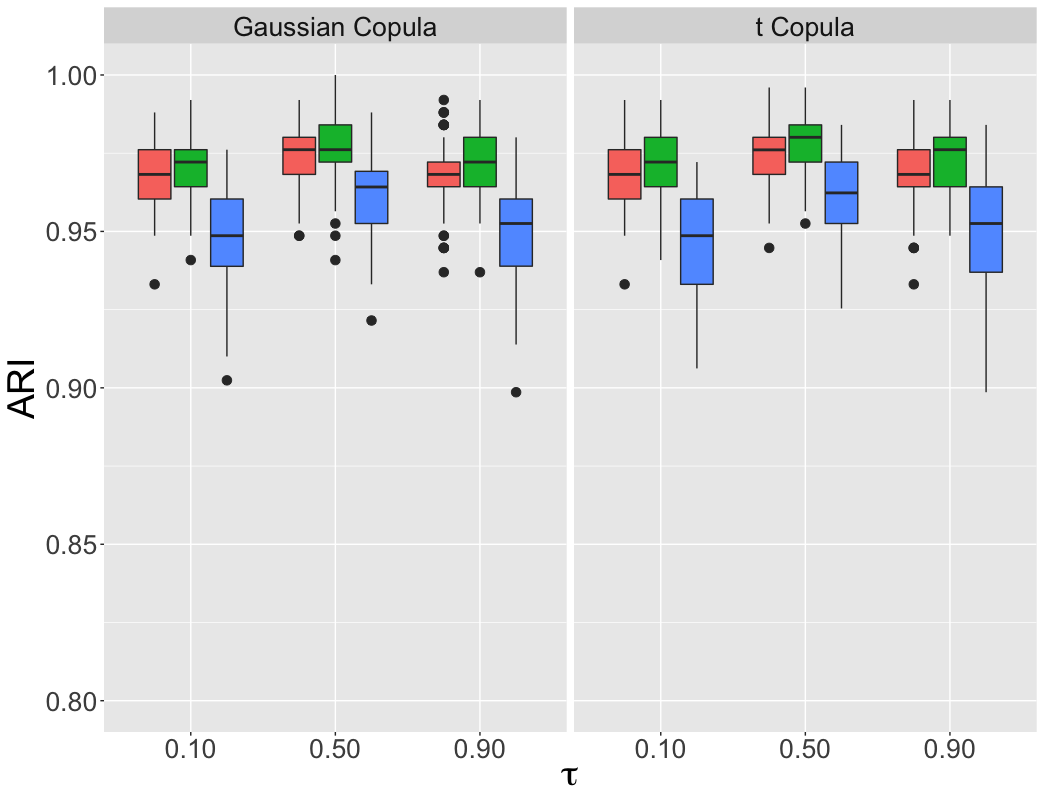}
\end{subfigure}
\hfill
\begin{subfigure}[b]{1.0\textwidth}
\includegraphics[width=.5\linewidth]{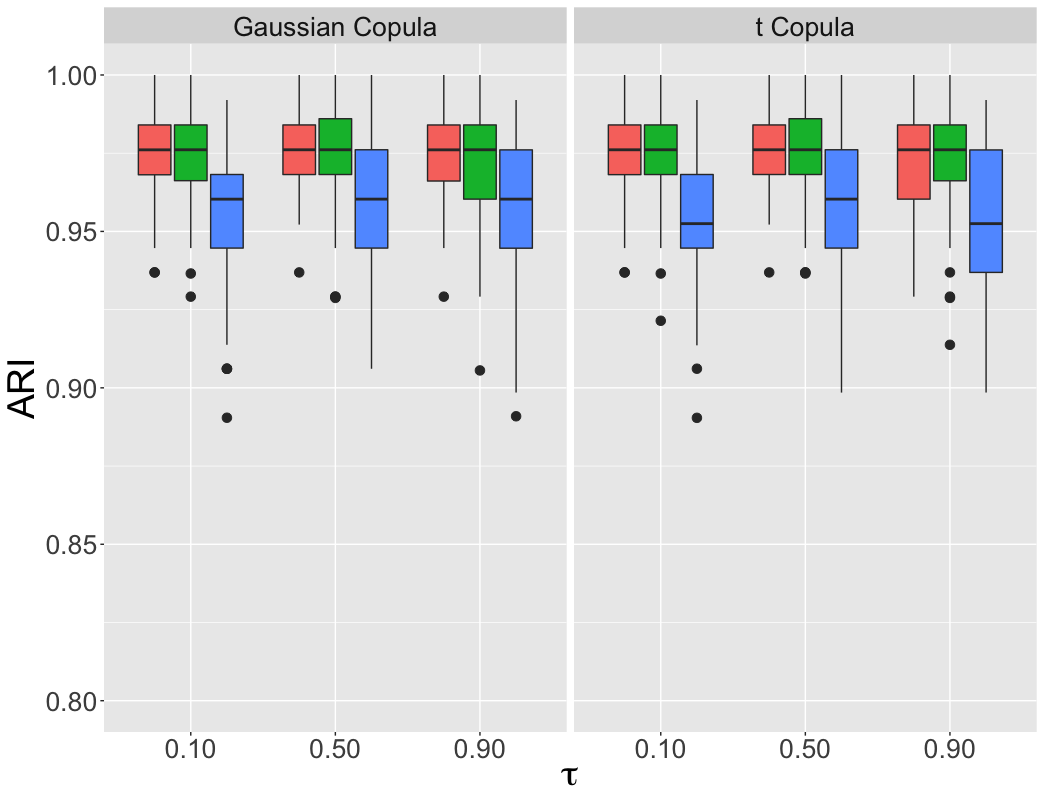}
\includegraphics[width=.5\linewidth]{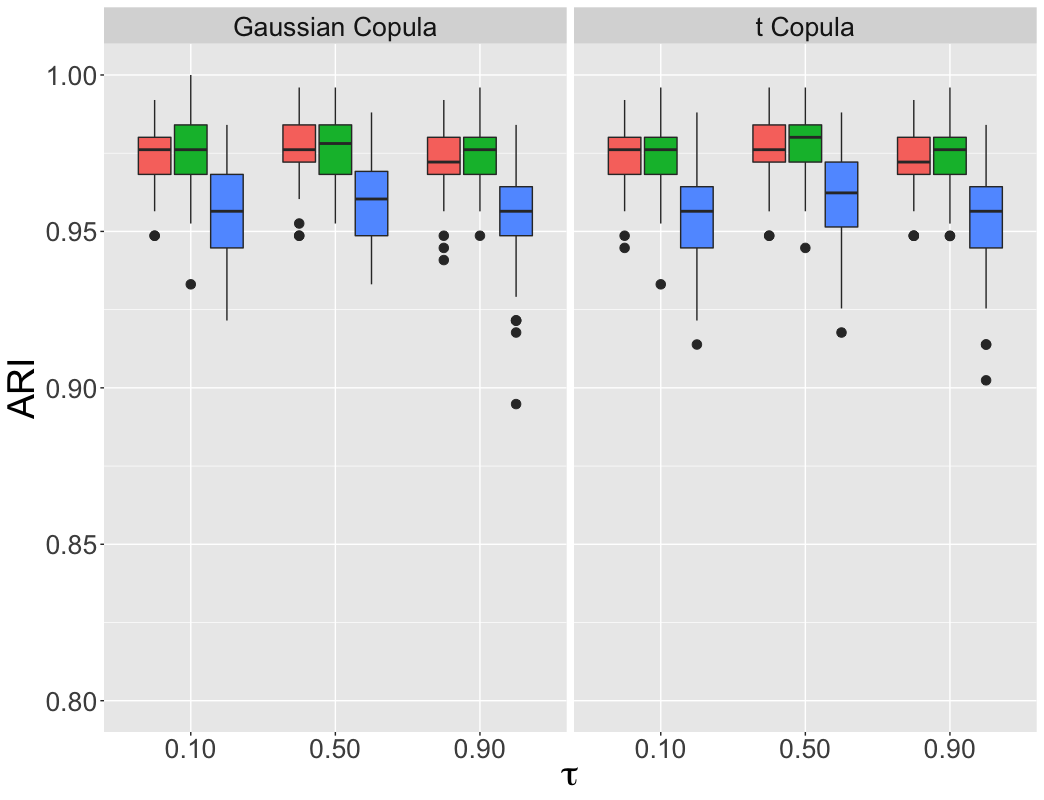}
\end{subfigure}
\caption{Box-plots of ARI for the posterior probabilities for CQHMM (first row) and CEHMM (second row) under the Gaussian (red), Student's t (green) and skew-t (blue) distributed errors with Gaussian and t copula, and sample sizes $T=500$ (left column) and $T=1000$ (right column).}
\label{fig:ari}
\end{figure}

 In order to evaluate the ability in recovering the true states partition we consider the Adjusted Rand Index (ARI) of \cite{hubert1985comparing}. The state partition provided by the fitted models is obtained by taking the maximum a posteriori probability, $\underset{k}{\max} \, \gamma_t (k)$, for every $t = 1,\dots,T$. In Figure \ref{fig:ari} is reported the box-plot of ARI for the posterior probabilities for all the settings considered for $T=500$ and $T = 1000$ sample sizes. For the CQHMM, we obtain very high performances in estimating the true state partition. For both copula settings, we obtain a better clustering performance for the model with Gaussian and t errors with respect to the skew-\textit{t} case. We also highlight that the cluster recovering ability depends on the specific quantile level, being the values slightly higher at the median ($\tau = 0.50$) than at the tails. Finally, when increasing the sample size to $T = 1000$, results slightly improve, reporting a lower variability for both error distributions.  
 Moving onto the CEHMM, we observe that in every setting we obtain very high performances in estimating the true state partition. For both copula settings, we obtain a better clustering performance for the model with Gaussian and t errors with respect to the skew-\textit{t} case. Compare to the CQHMM, the goodness of the clustering obtained does not depend on the specific expectile level, being the values very alike along the centre and the tails of the distribution, probably due to the the major efficiency of the algorithm in the expectile case. Finally, when increasing the sample size to $T = 1000$, results slightly improve reporting a lower variability for all error distributions. As in the quantile framework, the proposed CEHMM is able to recover the true values of the parameters and the true state partition in a highly satisfactory way in all the distributions and copula settings examined.

\begin{table}[htbp]
  \centering
  \scalebox{0.55}{
\begin{tabular}{rlrrrrlrrr}
\multicolumn{1}{l}{\textbf{CQHMM}} & \multicolumn{1}{r}{$\tau$} & 0.10  & 0.50  & 0.90  &       & \multicolumn{1}{r}{$\tau$} & 0.10  & 0.50  & 0.90 \\
\midrule
\multicolumn{1}{l}{Gaussian Copula} &       & Bias (Std.Err) & Bias (Std.Err) & Bias (Std.Err) & \multicolumn{1}{l}{t Copula} &       & Bias (Std.Err) & Bias (Std.Err) & Bias (Std.Err) \\
\multicolumn{1}{l}{Panel A: T=500} &       &       &       &       & \multicolumn{1}{l}{Panel A: T=500} &       &       &       &  \\
\multicolumn{1}{l}{State 1} &       &       &       &       & \multicolumn{1}{l}{State 1} &       &       &       &  \\
\multicolumn{1}{l}{j=1} & $\beta_{0,1}$ = -2 & -0.002 (0.076) & 0.004 (0.058) & 0.013 (0.079) & \multicolumn{1}{l}{j=1} & $\beta_{0,1}$ = -2 & -0.001 (0.083) & 0.005 (0.057) & 0.012 (0.082) \\
      & $\beta_{1,1}$ = 1 & -0.013 (0.130) & -0.008 (0.078) & 0.002 (0.123) &       & $\beta_{1,1}$ = 1 & -0.014 (0.133) & -0.009 (0.077) & -0.001 (0.123) \\
\multicolumn{1}{l}{j=2} & $\beta_{0,1}$ = 3 & -0.013 (0.088) & 0.000 (0.055) & 0.009 (0.086) & \multicolumn{1}{l}{j=2} & $\beta_{0,1}$ = 3 & -0.057 (0.455) & 0.000 (0.055) & 0.009 (0.087) \\
      & $\beta_{1,1}$ = -2 & 0.025 (0.108) & 0.008 (0.079) & -0.009 (0.105) &       & $\beta_{1,1}$ = -2 & 0.054 (0.300) & 0.008 (0.079) & -0.006 (0.106) \\
\multicolumn{1}{l}{State 2} &       &       &       &       & \multicolumn{1}{l}{State 2} &       &       &       &  \\
\multicolumn{1}{l}{j=1} & $\beta_{0,2}$ = 3 & -0.008 (0.076) & 0.001 (0.056) & 0.000 (0.077) & \multicolumn{1}{l}{j=1} & $\beta_{0,2}$ = 3 & -0.020 (0.123) & 0.000 (0.056) & -0.001 (0.078) \\
      & $\beta_{1,2}$ = -2 & -0.002 (0.127) & 0.003 (0.087) & 0.017 (0.107) &       & $\beta_{1,2}$ = -2 & 0.006 (0.176) & 0.004 (0.088) & 0.017 (0.106) \\
\multicolumn{1}{l}{j=2} & $\beta_{0,2}$ = -2 & 0.003 (0.073) & -0.005 (0.050) & -0.007 (0.093) & \multicolumn{1}{l}{j=2} & $\beta_{0,2}$ = -2 & -0.005 (0.096) & -0.005 (0.051) & -0.005 (0.094) \\
      & $\beta_{1,2}$ = 1 & -0.011 (0.110) & 0.013 (0.082) & 0.008 (0.123) &       & $\beta_{1,2}$ = 1 & 0.015 (0.300) & 0.014 (0.083) & 0.010 (0.122) \\
      &       &       &       &       &       &       &       &       &  \\
\multicolumn{1}{l}{Panel B: T=1000} &       &       &       &       & \multicolumn{1}{l}{Panel B: T=1000} &       &       &       &  \\
\multicolumn{1}{l}{State 1} &       &       &       &       & \multicolumn{1}{l}{State 1} &       &       &       &  \\
\multicolumn{1}{l}{j=1} & $\beta_{0,1}$ = -2 & -0.005 (0.064) & -0.005 (0.042) & -0.017 (0.054) & \multicolumn{1}{l}{j=1} & $\beta_{0,1}$ = -2 & -0.005 (0.064) & -0.004 (0.042) & -0.014 (0.054) \\
      & $\beta_{1,1}$ = 1 & -0.013 (0.080) & -0.015 (0.045) & -0.002 (0.071) &       & $\beta_{1,1}$ = 1 & -0.013 (0.081) & -0.015 (0.045) & -0.002 (0.070) \\
\multicolumn{1}{l}{j=2} & $\beta_{0,1}$ = 3 & -0.008 (0.053) & -0.002 (0.038) & 0.007 (0.066) & \multicolumn{1}{l}{j=2} & $\beta_{0,1}$ = 3 & -0.009 (0.052) & -0.001 (0.038) & 0.007 (0.066) \\
      & $\beta_{1,1}$ = -2 & 0.001 (0.075) & 0.006 (0.056) & 0.013 (0.076) &       & $\beta_{1,1}$ = -2 & 0.002 (0.073) & 0.007 (0.056) & 0.015 (0.076) \\
\multicolumn{1}{l}{State 2} &       &       &       &       & \multicolumn{1}{l}{State 2} &       &       &       &  \\
\multicolumn{1}{l}{j=1} & $\beta_{0,2}$ = 3 & 0.007 (0.072) & 0.003 (0.042) & 0.018 (0.063) & \multicolumn{1}{l}{j=1} & $\beta_{0,2}$ = 3 & 0.002 (0.071) & 0.003 (0.042) & 0.017 (0.064) \\
      & $\beta_{1,2}$ = -2 & -0.001 (0.081) & 0.007 (0.056) & 0.005 (0.080) &       & $\beta_{1,2}$ = -2 & -0.005 (0.081) & 0.006 (0.057) & 0.004 (0.080) \\
\multicolumn{1}{l}{j=2} & $\beta_{0,2}$ = -2 & -0.006 (0.050) & 0.000 (0.039) & -0.002 (0.062) & \multicolumn{1}{l}{j=2} & $\beta_{0,2}$ = -2 & -0.006 (0.050) & 0.001 (0.040) & 0.000 (0.062) \\
      & $\beta_{1,2}$ = 1 & -0.003 (0.075) & 0.000 (0.053) & -0.003 (0.077) &       & $\beta_{1,2}$ = 1 & -0.004 (0.074) & 0.001 (0.053) & 0.000 (0.077) \\
\bottomrule
\end{tabular}}%
      \caption{Bias and standard error values of the state-regression parameter estimates for CQHMM with Gaussian distributed errors for $T = 500$ (Panel A) and $T = 1000$ (Panel B).}
  \label{tab:sim_ndist_q}%
\end{table}%

\begin{table}[htbp]
  \centering
  \scalebox{0.55}{
\begin{tabular}{rlrrrrlrrr}
\multicolumn{1}{l}{\textbf{CQHMM}} & \multicolumn{1}{r}{$\tau$} & 0.10  & 0.50  & 0.90  &       & \multicolumn{1}{r}{$\tau$} & 0.10  & 0.50  & 0.90 \\
\midrule
\multicolumn{1}{l}{Gaussian Copula} &       & Bias (Std.Err) & Bias (Std.Err) & Bias (Std.Err) & \multicolumn{1}{l}{t Copula} &       & Bias (Std.Err) & Bias (Std.Err) & Bias (Std.Err) \\
\multicolumn{1}{l}{Panel A: T=500} &       &       &       &       & \multicolumn{1}{l}{Panel A: T=500} &       &       &       &  \\
\multicolumn{1}{l}{State 1} &       &       &       &       & \multicolumn{1}{l}{State 1} &       &       &       &  \\
\multicolumn{1}{l}{j=1} & $\beta_{0,1}$ = -2 & -0.066 (0.405) & -0.003 (0.069) & 0.283 (1.004) & \multicolumn{1}{l}{j=1} & $\beta_{0,1}$ = -2 & -0.116 (0.715) & -0.003 (0.069) & 0.360 (1.186) \\
      & $\beta_{1,1}$ = 1 & -0.026 (0.239) & 0.011 (0.094) & -0.202 (1.152) &       & $\beta_{1,1}$ = 1 & -0.097 (0.799) & 0.011 (0.094) & -0.248 (0.869) \\
\multicolumn{1}{l}{j=2} & $\beta_{0,1}$ = 3 & -0.297 (1.155) & -0.001 (0.067) & -0.145 (0.884) & \multicolumn{1}{l}{j=2} & $\beta_{0,1}$ = 3 & -0.455 (1.438) & 0.000 (0.068) & -0.094 (0.585) \\
      & $\beta_{1,1}$ = -2 & 0.176 (1.021) & -0.019 (0.079) & -0.108 (1.279) &       & $\beta_{1,1}$ = -2 & 0.420 (1.331) & -0.018 (0.080) & 0.012 (0.343) \\
\multicolumn{1}{l}{State 2} &       &       &       &       & \multicolumn{1}{l}{State 2} &       &       &       &  \\
\multicolumn{1}{l}{j=1} & $\beta_{0,2}$ = 3 & -0.343 (1.153) & 0.010 (0.070) & 0.013 (0.215) & \multicolumn{1}{l}{j=1} & $\beta_{0,2}$ = 3 & -0.497 (1.331) & 0.011 (0.071) & 0.071 (0.343) \\
      & $\beta_{1,2}$ = -2 & 0.185 (1.001) & 0.002 (0.079) & 0.005 (0.270) &       & $\beta_{1,2}$ = -2 & 0.405 (1.268) & 0.004 (0.079) & 0.018 (0.191) \\
\multicolumn{1}{l}{j=2} & $\beta_{0,2}$ = -2 & 0.042 (0.701) & -0.003 (0.066) & 0.537 (1.553) & \multicolumn{1}{l}{j=2} & $\beta_{0,2}$ = -2 & 0.110 (0.822) & -0.002 (0.065) & 0.446 (1.464) \\
      & $\beta_{1,2}$ = 1 & -0.001 (0.549) & 0.004 (0.082) & -0.334 (1.034) &       & $\beta_{1,2}$ = 1 & -0.101 (0.584) & 0.004 (0.082) & -0.199 (0.781) \\
      &       &       &       &       &       &       &       &       &  \\
\multicolumn{1}{l}{Panel B: T=1000} &       &       &       &       & \multicolumn{1}{l}{Panel B: T=1000} &       &       &       &  \\
\multicolumn{1}{l}{State 1} &       &       &       &       & \multicolumn{1}{l}{State 1} &       &       &       &  \\
\multicolumn{1}{l}{j=1} & $\beta_{0,1}$ = -2 & -0.171 (0.727) & -0.002 (0.047) & 0.270 (1.083) & \multicolumn{1}{l}{j=1} & $\beta_{0,1}$ = -2 & -0.133 (0.482) & -0.001 (0.047) & 0.541 (1.396) \\
      & $\beta_{1,1}$ = 1 & 0.068 (1.074) & -0.002 (0.065) & -0.330 (1.309) &       & $\beta_{1,1}$ = 1 & -0.065 (0.268) & -0.002 (0.066) & -0.479 (1.225) \\
\multicolumn{1}{l}{j=2} & $\beta_{0,1}$ = 3 & -0.563 (1.560) & 0.006 (0.043) & -0.006 (0.718) & \multicolumn{1}{l}{j=2} & $\beta_{0,1}$ = 3 & -1.006 (1.850) & 0.006 (0.044) & 0.022 (0.274) \\
      & $\beta_{1,1}$ = -2 & 0.576 (1.705) & -0.007 (0.065) & 0.046 (0.963) &       & $\beta_{1,1}$ = -2 & 0.716 (1.314) & -0.008 (0.066) & -0.015 (0.332) \\
\multicolumn{1}{l}{State 2} &       &       &       &       & \multicolumn{1}{l}{State 2} &       &       &       &  \\
\multicolumn{1}{l}{j=1} & $\beta_{0,2}$ = 3 & -0.412 (1.377) & 0.002 (0.044) & 0.044 (0.270) & \multicolumn{1}{l}{j=1} & $\beta_{0,2}$ = 3 & -0.936 (1.711) & 0.001 (0.044) & 0.083 (0.292) \\
      & $\beta_{1,2}$ = -2 & 0.554 (1.943) & 0.002 (0.061) & 0.004 (0.312) &       & $\beta_{1,2}$ = -2 & 0.662 (1.228) & 0.002 (0.061) & 0.042 (0.170) \\
\multicolumn{1}{l}{j=2} & $\beta_{0,2}$ = -2 & 0.184 (1.299) & -0.009 (0.041) & 0.421 (1.327) & \multicolumn{1}{l}{j=2} & $\beta_{0,2}$ = -2 & 0.002 (0.243) & -0.006 (0.041) & 0.616 (1.500) \\
      & $\beta_{1,2}$ = 1 & 0.059 (2.221) & -0.004 (0.061) & -0.269 (1.005) &       & $\beta_{1,2}$ = 1 & -0.049 (0.151) & -0.002 (0.062) & -0.370 (1.003) \\
\bottomrule
\end{tabular}%
}%
        \caption{Bias and standard error values of the state-regression parameter estimates for CQHMM with Student's t distributed errors for $T = 500$ (Panel A) and $T = 1000$ (Panel B).}
  \label{tab:sim_tdist_q}%
\end{table}%

\begin{table}[htbp]
  \centering
  \scalebox{0.55}{
\begin{tabular}{rlrrrrlrrr}
\multicolumn{1}{l}{\textbf{CQHMM}} & \multicolumn{1}{r}{$\tau$} & 0.10  & 0.50  & 0.90  &       & \multicolumn{1}{r}{$\tau$} & 0.10  & 0.50  & 0.90 \\
\midrule
\multicolumn{1}{l}{Gaussian Copula} &       & Bias (Std.Err) & Bias (Std.Err) & Bias (Std.Err) & \multicolumn{1}{l}{Student's t Copula} &       & Bias (Std.Err) & Bias (Std.Err) & Bias (Std.Err) \\
\multicolumn{1}{l}{Panel A: T=500} &       &       &       &       & \multicolumn{1}{l}{Panel A: T=500} &       &       &       &  \\
\multicolumn{1}{l}{State 1} &       &       &       &       & \multicolumn{1}{l}{State 1} &       &       &       &  \\
\multicolumn{1}{l}{j=1} & $\beta_{0,1}$ = -2 & -0.117 (0.220) & -0.100 (0.057) & -1.047 (10.412) & \multicolumn{1}{l}{j=1} & $\beta_{0,1}$ = -2 & -0.117 (0.277) & -0.099 (0.059) & -0.038 (0.786) \\
      & $\beta_{1,1}$ = 1 & -0.041 (0.185) & 0.010 (0.073) & 1.189 (13.284) &       & $\beta_{1,1}$ = 1 & -0.027 (0.239) & 0.011 (0.074) & -0.087 (0.564) \\
\multicolumn{1}{l}{j=2} & $\beta_{0,1}$ = 3 & -0.253 (1.524) & 0.099 (0.060) & 0.472 (4.430) & \multicolumn{1}{l}{j=2} & $\beta_{0,1}$ = 3 & -0.314 (1.557) & 0.099 (0.060) & 0.035 (0.253) \\
      & $\beta_{1,1}$ = -2 & 0.206 (0.777) & 0.007 (0.084) & -0.606 (6.110) &       & $\beta_{1,1}$ = -2 & 0.258 (0.872) & 0.008 (0.083) & -0.023 (0.347) \\
\multicolumn{1}{l}{State 2} &       &       &       &       & \multicolumn{1}{l}{State 2} &       &       &       &  \\
\multicolumn{1}{l}{j=1} & $\beta_{0,2}$ = 3 & -0.266 (1.189) & 0.122 (0.070) & 0.215 (0.281) & \multicolumn{1}{l}{j=1} & $\beta_{0,2}$ = 3 & -0.305 (1.230) & 0.119 (0.068) & 0.186 (0.148) \\
      & $\beta_{1,2}$ = -2 & 0.237 (0.853) & 0.000 (0.067) & 0.009 (0.152) &       & $\beta_{1,2}$ = -2 & 0.291 (1.072) & -0.002 (0.066) & 0.016 (0.159) \\
\multicolumn{1}{l}{j=2} & $\beta_{0,2}$ = -2 & 0.056 (1.139) & -0.111 (0.065) & 0.291 (1.418) & \multicolumn{1}{l}{j=2} & $\beta_{0,2}$ = -2 & -0.010 (0.919) & -0.110 (0.065) & 0.349 (1.424) \\
      & $\beta_{1,2}$ = 1 & -0.119 (0.616) & -0.001 (0.080) & -0.189 (0.683) &       & $\beta_{1,2}$ = 1 & -0.111 (0.566) & 0.000 (0.081) & -0.212 (0.733) \\
      &       &       &       &       &       &       &       &       &  \\
\multicolumn{1}{l}{Panel B: T=1000} &       &       &       &       & \multicolumn{1}{l}{Panel B: T=1000} &       &       &       &  \\
\multicolumn{1}{l}{State 1} &       &       &       &       & \multicolumn{1}{l}{State 1} &       &       &       &  \\
\multicolumn{1}{l}{j=1} & $\beta_{0,1}$ = -2 & -0.152 (0.487) & -0.101 (0.044) & -0.018 (1.094) & \multicolumn{1}{l}{j=1} & $\beta_{0,1}$ = -2 & -0.141 (0.301) & -0.101 (0.045) & 0.168 (1.240) \\
      & $\beta_{1,1}$ = 1 & 0.070 (0.686) & -0.003 (0.056) & -0.266 (1.217) &       & $\beta_{1,1}$ = 1 & -0.009 (0.143) & -0.002 (0.056) & -0.244 (0.795) \\
\multicolumn{1}{l}{j=2} & $\beta_{0,1}$ = 3 & -0.274 (1.476) & 0.103 (0.044) & 0.167 (0.737) & \multicolumn{1}{l}{j=2} & $\beta_{0,1}$ = 3 & -0.318 (1.568) & 0.103 (0.044) & 0.051 (0.184) \\
      & $\beta_{1,1}$ = -2 & 0.285 (0.988) & -0.011 (0.048) & 0.099 (0.909) &       & $\beta_{1,1}$ = -2 & 0.298 (0.904) & -0.011 (0.048) & 0.001 (0.126) \\
\multicolumn{1}{l}{State 2} &       &       &       &       & \multicolumn{1}{l}{State 2} &       &       &       &  \\
\multicolumn{1}{l}{j=1} & $\beta_{0,2}$ = 3 & -0.308 (1.239) & 0.112 (0.046) & 0.238 (0.410) & \multicolumn{1}{l}{j=1} & $\beta_{0,2}$ = 3 & -0.344 (1.281) & 0.111 (0.045) & 0.212 (0.221) \\
      & $\beta_{1,2}$ = -2 & 0.348 (1.138) & 0.006 (0.060) & 0.012 (0.275) &       & $\beta_{1,2}$ = -2 & 0.284 (0.875) & 0.004 (0.059) & 0.092 (0.691) \\
\multicolumn{1}{l}{j=2} & $\beta_{0,2}$ = -2 & -0.099 (0.293) & -0.119 (0.047) & 0.205 (1.117) & \multicolumn{1}{l}{j=2} & $\beta_{0,2}$ = -2 & -0.112 (0.589) & -0.118 (0.046) & 0.385 (1.476) \\
      & $\beta_{1,2}$ = 1 & -0.056 (0.671) & 0.000 (0.065) & -0.244 (1.071) &       & $\beta_{1,2}$ = 1 & -0.056 (0.308) & 0.001 (0.064) & -0.268 (0.932) \\
\bottomrule
\end{tabular}}%
        \caption{Bias and standard error values of the state-regression parameter estimates for CQHMM with skew t distributed errors for $T = 500$ (Panel A) and $T = 1000$ (Panel B).}
  \label{tab:sim_stdist_q}%
\end{table}%

\begin{table}[htbp]
  \centering
  \scalebox{0.55}{
\begin{tabular}{rlrrrrlrrr}
\multicolumn{1}{l}{\textbf{CEHMM}} & \multicolumn{1}{r}{$\tau$} & 0.10  & 0.50  & 0.90  &       & \multicolumn{1}{r}{$\tau$} & 0.10  & 0.50  & 0.90 \\
\midrule
\multicolumn{1}{l}{Gaussian Copula} &       & Bias (Std.Err) & Bias (Std.Err) & Bias (Std.Err) & \multicolumn{1}{l}{Student's t Copula} &       & Bias (Std.Err) & Bias (Std.Err) & Bias (Std.Err) \\
\multicolumn{1}{l}{Panel A: T=500} &       &       &       &       & \multicolumn{1}{l}{Panel A: T=500} &       &       &       &  \\
\multicolumn{1}{l}{State 1} &       &       &       &       & \multicolumn{1}{l}{State 1} &       &       &       &  \\
\multicolumn{1}{l}{j=1} & $\beta_{0,1}$ = -2 & 0.000 (0.075) & 0.000 (0.060) & -0.009 (0.081) & \multicolumn{1}{l}{j=1} & $\beta_{0,1}$ = -2 & 0.003 (0.075) & 0.001 (0.060) & -0.008 (0.081) \\
      & $\beta_{1,1}$ = 1 & -0.009 (0.093) & -0.009 (0.070) & -0.007 (0.087) &       & $\beta_{1,1}$ = 1 & -0.009 (0.092) & -0.009 (0.070) & -0.007 (0.087) \\
\multicolumn{1}{l}{j=2} & $\beta_{0,1}$ = 3 & 0.001 (0.072) & 0.000 (0.057) & -0.002 (0.072) & \multicolumn{1}{l}{j=2} & $\beta_{0,1}$ = 3 & 0.004 (0.072) & 0.001 (0.057) & -0.001 (0.071) \\
      & $\beta_{1,1}$ = -2 & 0.012 (0.074) & 0.005 (0.058) & -0.005 (0.073) &       & $\beta_{1,1}$ = -2 & 0.012 (0.074) & 0.005 (0.058) & -0.004 (0.073) \\
\multicolumn{1}{l}{State 2} &       &       &       &       & \multicolumn{1}{l}{State 2} &       &       &       &  \\
\multicolumn{1}{l}{j=1} & $\beta_{0,2}$ = 3 & -0.004 (0.067) & -0.005 (0.052) & -0.013 (0.072) & \multicolumn{1}{l}{j=1} & $\beta_{0,2}$ = 3 & -0.005 (0.067) & -0.005 (0.052) & -0.011 (0.071) \\
      & $\beta_{1,2}$ = -2 & -0.004 (0.084) & 0.003 (0.071) & 0.008 (0.084) &       & $\beta_{1,2}$ = -2 & -0.005 (0.084) & 0.003 (0.071) & 0.008 (0.084) \\
\multicolumn{1}{l}{j=2} & $\beta_{0,2}$ = -2 & 0.008 (0.063) & -0.003 (0.053) & -0.012 (0.079) & \multicolumn{1}{l}{j=2} & $\beta_{0,2}$ = -2 & 0.007 (0.063) & -0.003 (0.053) & -0.012 (0.079) \\
      & $\beta_{1,2}$ = 1 & 0.001 (0.083) & 0.008 (0.067) & 0.012 (0.085) &       & $\beta_{1,2}$ = 1 & 0.001 (0.083) & 0.008 (0.067) & 0.012 (0.085) \\
      &       &       &       &       &       &       &       &       &  \\
\multicolumn{1}{l}{Panel B: T=1000} &       &       &       &       & \multicolumn{1}{l}{Panel B: T=1000} &       &       &       &  \\
\multicolumn{1}{l}{State 1} &       &       &       &       & \multicolumn{1}{l}{State 1} &       &       &       &  \\
\multicolumn{1}{l}{j=1} & $\beta_{0,1}$ = -2 & -0.004 (0.057) & -0.008 (0.040) & -0.018 (0.055) & \multicolumn{1}{l}{j=1} & $\beta_{0,1}$ = -2 & -0.001 (0.057) & -0.005 (0.040) & -0.014 (0.055) \\
      & $\beta_{1,1}$ = 1 & -0.007 (0.059) & -0.010 (0.042) & -0.009 (0.050) &       & $\beta_{1,1}$ = 1 & -0.007 (0.059) & -0.010 (0.042) & -0.009 (0.050) \\
\multicolumn{1}{l}{j=2} & $\beta_{0,1}$ = 3 & 0.010 (0.049) & 0.006 (0.040) & 0.002 (0.054) & \multicolumn{1}{l}{j=2} & $\beta_{0,1}$ = 3 & 0.010 (0.049) & 0.006 (0.040) & 0.004 (0.054) \\
      & $\beta_{1,1}$ = -2 & 0.006 (0.049) & 0.006 (0.043) & 0.007 (0.053) &       & $\beta_{1,1}$ = -2 & 0.006 (0.049) & 0.006 (0.043) & 0.007 (0.053) \\
\multicolumn{1}{l}{State 2} &       &       &       &       & \multicolumn{1}{l}{State 2} &       &       &       &  \\
\multicolumn{1}{l}{j=1} & $\beta_{0,2}$ = 3 & 0.004 (0.059) & 0.005 (0.048) & 0.001 (0.061) & \multicolumn{1}{l}{j=1} & $\beta_{0,2}$ = 3 & -0.002 (0.059) & 0.001 (0.048) & -0.002 (0.061) \\
      & $\beta_{1,2}$ = -2 & -0.002 (0.060) & 0.004 (0.048) & 0.007 (0.056) &       & $\beta_{1,2}$ = -2 & -0.003 (0.060) & 0.003 (0.048) & 0.007 (0.056) \\
\multicolumn{1}{l}{j=2} & $\beta_{0,2}$ = -2 & 0.008 (0.046) & 0.004 (0.041) & -0.007 (0.059) & \multicolumn{1}{l}{j=2} & $\beta_{0,2}$ = -2 & 0.008 (0.047) & 0.003 (0.041) & -0.008 (0.059) \\
      & $\beta_{1,2}$ = 1 & -0.007 (0.052) & -0.004 (0.041) & -0.003 (0.051) &       & $\beta_{1,2}$ = 1 & -0.007 (0.052) & -0.004 (0.041) & -0.003 (0.051) \\
\bottomrule
\end{tabular}%
}%
      \caption{Bias and standard error values of the state-regression parameter estimates for CEHMM with Gaussian distributed errors for $T = 500$ (Panel A) and $T = 1000$ (Panel B).}
  \label{tab:sim_ndist_e}%
\end{table}%

\begin{table}[htbp]
  \centering
  \scalebox{0.55}{
\begin{tabular}{rlrrrrlrrr}
\multicolumn{1}{l}{\textbf{CEHMM}} & \multicolumn{1}{r}{$\tau$} & 0.10  & 0.50  & 0.90  &       & \multicolumn{1}{r}{$\tau$} & 0.10  & 0.50  & 0.90 \\
\midrule
\multicolumn{1}{l}{Gaussian Copula} &       & Bias (Std.Err) & Bias (Std.Err) & Bias (Std.Err) & \multicolumn{1}{l}{Student's t Copula} &       & Bias (Std.Err) & Bias (Std.Err) & Bias (Std.Err) \\
\multicolumn{1}{l}{Panel A: T=500} &       &       &       &       & \multicolumn{1}{l}{Panel A: T=500} &       &       &       &  \\
\multicolumn{1}{l}{State 1} &       &       &       &       & \multicolumn{1}{l}{State 1} &       &       &       &  \\
\multicolumn{1}{l}{j=1} & $\beta_{0,1}$ = -2 & 0.002 (0.138) & -0.002 (0.091) & -0.014 (0.152) & \multicolumn{1}{l}{j=1} & $\beta_{0,1}$ = -2 & 0.003 (0.137) & 0.001 (0.091) & -0.004 (0.156) \\
      & $\beta_{1,1}$ = 1 & 0.000 (0.116) & 0.010 (0.082) & 0.013 (0.141) &       & $\beta_{1,1}$ = 1 & -0.005 (0.119) & 0.009 (0.084) & 0.018 (0.144) \\
\multicolumn{1}{l}{j=2} & $\beta_{0,1}$ = 3 & 0.023 (0.154) & 0.006 (0.094) & -0.009 (0.143) & \multicolumn{1}{l}{j=2} & $\beta_{0,1}$ = 3 & 0.023 (0.155) & 0.007 (0.094) & -0.002 (0.146) \\
      & $\beta_{1,1}$ = -2 & -0.007 (0.122) & -0.009 (0.079) & -0.001 (0.125) &       & $\beta_{1,1}$ = -2 & -0.018 (0.130) & -0.010 (0.080) & 0.002 (0.125) \\
\multicolumn{1}{l}{State 2} &       &       &       &       & \multicolumn{1}{l}{State 2} &       &       &       &  \\
\multicolumn{1}{l}{j=1} & $\beta_{0,2}$ = 3 & 0.014 (0.124) & 0.007 (0.075) & 0.000 (0.122) & \multicolumn{1}{l}{j=1} & $\beta_{0,2}$ = 3 & 0.012 (0.129) & 0.007 (0.075) & -0.001 (0.121) \\
      & $\beta_{1,2}$ = -2 & -0.003 (0.128) & -0.002 (0.076) & 0.003 (0.141) &       & $\beta_{1,2}$ = -2 & 0.001 (0.131) & -0.002 (0.077) & -0.001 (0.141) \\
\multicolumn{1}{l}{j=2} & $\beta_{0,2}$ = -2 & -0.007 (0.146) & -0.002 (0.084) & 0.001 (0.155) & \multicolumn{1}{l}{j=2} & $\beta_{0,2}$ = -2 & -0.002 (0.139) & -0.001 (0.084) & 0.003 (0.146) \\
      & $\beta_{1,2}$ = 1 & 0.007 (0.127) & 0.005 (0.072) & 0.007 (0.118) &       & $\beta_{1,2}$ = 1 & 0.015 (0.123) & 0.007 (0.072) & 0.007 (0.112) \\
      &       &       &       &       &       &       &       &       &  \\
\multicolumn{1}{l}{Panel B: T=1000} &       &       &       &       & \multicolumn{1}{l}{Panel B: T=1000} &       &       &       &  \\
\multicolumn{1}{l}{State 1} &       &       &       &       & \multicolumn{1}{l}{State 1} &       &       &       &  \\
\multicolumn{1}{l}{j=1} & $\beta_{0,1}$ = -2 & 0.006 (0.091) & -0.006 (0.060) & -0.033 (0.102) & \multicolumn{1}{l}{j=1} & $\beta_{0,1}$ = -2 & 0.000 (0.094) & -0.008 (0.060) & -0.024 (0.117) \\
      & $\beta_{1,1}$ = 1 & 0.004 (0.093) & -0.004 (0.058) & -0.019 (0.098) &       & $\beta_{1,1}$ = 1 & 0.000 (0.093) & -0.003 (0.058) & -0.012 (0.102) \\
\multicolumn{1}{l}{j=2} & $\beta_{0,1}$ = 3 & 0.035 (0.093) & 0.021 (0.058) & 0.020 (0.095) & \multicolumn{1}{l}{j=2} & $\beta_{0,1}$ = 3 & 0.029 (0.099) & 0.016 (0.058) & 0.013 (0.097) \\
      & $\beta_{1,1}$ = -2 & -0.001 (0.089) & -0.009 (0.060) & -0.012 (0.100) &       & $\beta_{1,1}$ = -2 & -0.005 (0.094) & -0.011 (0.060) & -0.014 (0.095) \\
\multicolumn{1}{l}{State 2} &       &       &       &       & \multicolumn{1}{l}{State 2} &       &       &       &  \\
\multicolumn{1}{l}{j=1} & $\beta_{0,2}$ = 3 & 0.009 (0.099) & 0.005 (0.057) & 0.016 (0.095) & \multicolumn{1}{l}{j=1} & $\beta_{0,2}$ = 3 & 0.003 (0.100) & 0.001 (0.057) & 0.007 (0.098) \\
      & $\beta_{1,2}$ = -2 & -0.001 (0.090) & 0.002 (0.059) & 0.010 (0.102) &       & $\beta_{1,2}$ = -2 & 0.000 (0.086) & 0.001 (0.060) & 0.007 (0.098) \\
\multicolumn{1}{l}{j=2} & $\beta_{0,2}$ = -2 & 0.010 (0.102) & 0.003 (0.056) & 0.012 (0.093) & \multicolumn{1}{l}{j=2} & $\beta_{0,2}$ = -2 & 0.012 (0.103) & 0.003 (0.057) & 0.015 (0.100) \\
      & $\beta_{1,2}$ = 1 & -0.002 (0.098) & 0.005 (0.059) & 0.013 (0.090) &       & $\beta_{1,2}$ = 1 & 0.000 (0.099) & 0.007 (0.059) & 0.017 (0.091) \\
\bottomrule
\end{tabular}%
}%
        \caption{Bias and standard error values of the state-regression parameter estimates for CEHMM with Student's t distributed errors for $T = 500$ (Panel A) and $T = 1000$ (Panel B).}
  \label{tab:sim_tdist_e}%
\end{table}%

\begin{table}[htbp]
  \centering
  \scalebox{0.55}{
\begin{tabular}{rlrrrrlrrr}
\multicolumn{1}{l}{\textbf{CEHMM}} & \multicolumn{1}{r}{$\tau$} & 0.10  & 0.50  & 0.90  &       & \multicolumn{1}{r}{$\tau$} & 0.10  & 0.50  & 0.90 \\
\midrule
\multicolumn{1}{l}{Gaussian Copula} &       & Bias (Std.Err) & Bias (Std.Err) & Bias (Std.Err) & \multicolumn{1}{l}{Student's t Copula} &       & Bias (Std.Err) & Bias (Std.Err) & Bias (Std.Err) \\
\multicolumn{1}{l}{Panel A: T=500} &       &       &       &       & \multicolumn{1}{l}{Panel A: T=500} &       &       &       &  \\
\multicolumn{1}{l}{State 1} &       &       &       &       & \multicolumn{1}{l}{State 1} &       &       &       &  \\
\multicolumn{1}{l}{j=1} & $\beta_{0,1}$ = -2 & 0.012 (0.152) & 0.008 (0.080) & 0.008 (0.112) & \multicolumn{1}{l}{j=1} & $\beta_{0,1}$ = -2 & 0.016 (0.151) & 0.009 (0.081) & 0.011 (0.109) \\
      & $\beta_{1,1}$ = 1 & -0.006 (0.130) & 0.006 (0.069) & 0.015 (0.102) &       & $\beta_{1,1}$ = 1 & -0.003 (0.128) & 0.007 (0.071) & 0.017 (0.101) \\
\multicolumn{1}{l}{j=2} & $\beta_{0,1}$ = 3 & -0.001 (0.106) & -0.007 (0.083) & -0.013 (0.153) & \multicolumn{1}{l}{j=2} & $\beta_{0,1}$ = 3 & -0.003 (0.108) & -0.005 (0.082) & -0.015 (0.149) \\
      & $\beta_{1,1}$ = -2 & -0.003 (0.099) & 0.002 (0.082) & 0.005 (0.155) &       & $\beta_{1,1}$ = -2 & -0.006 (0.104) & 0.002 (0.083) & 0.003 (0.148) \\
\multicolumn{1}{l}{State 2} &       &       &       &       & \multicolumn{1}{l}{State 2} &       &       &       &  \\
\multicolumn{1}{l}{j=1} & $\beta_{0,2}$ = 3 & 0.012 (0.132) & 0.010 (0.070) & -0.002 (0.102) & \multicolumn{1}{l}{j=1} & $\beta_{0,2}$ = 3 & 0.010 (0.141) & 0.011 (0.071) & 0.001 (0.102) \\
      & $\beta_{1,2}$ = -2 & -0.001 (0.143) & -0.001 (0.073) & 0.001 (0.120) &       & $\beta_{1,2}$ = -2 & -0.010 (0.144) & -0.003 (0.072) & 0.000 (0.119) \\
\multicolumn{1}{l}{j=2} & $\beta_{0,2}$ = -2 & 0.000 (0.117) & -0.002 (0.077) & 0.002 (0.151) & \multicolumn{1}{l}{j=2} & $\beta_{0,2}$ = -2 & 0.007 (0.118) & 0.001 (0.079) & 0.006 (0.151) \\
      & $\beta_{1,2}$ = 1 & 0.001 (0.113) & 0.002 (0.070) & 0.007 (0.135) &       & $\beta_{1,2}$ = 1 & 0.003 (0.113) & 0.002 (0.070) & 0.014 (0.132) \\
      &       &       &       &       &       &       &       &       &  \\
\multicolumn{1}{l}{Panel B: T=1000} &       &       &       &       & \multicolumn{1}{l}{Panel B: T=1000} &       &       &       &  \\
\multicolumn{1}{l}{State 1} &       &       &       &       & \multicolumn{1}{l}{State 1} &       &       &       &  \\
\multicolumn{1}{l}{j=1} & $\beta_{0,1}$ = -2 & 0.003 (0.100) & 0.003 (0.055) & 0.006 (0.075) & \multicolumn{1}{l}{j=1} & $\beta_{0,1}$ = -2 & 0.004 (0.101) & 0.004 (0.056) & 0.009 (0.082) \\
      & $\beta_{1,1}$ = 1 & 0.009 (0.100) & 0.003 (0.055) & 0.000 (0.078) &       & $\beta_{1,1}$ = 1 & 0.012 (0.099) & 0.003 (0.056) & 0.002 (0.079) \\
\multicolumn{1}{l}{j=2} & $\beta_{0,1}$ = 3 & -0.001 (0.070) & -0.001 (0.054) & -0.005 (0.105) & \multicolumn{1}{l}{j=2} & $\beta_{0,1}$ = 3 & 0.002 (0.069) & 0.001 (0.054) & -0.001 (0.108) \\
      & $\beta_{1,1}$ = -2 & -0.006 (0.065) & -0.010 (0.049) & -0.010 (0.109) &       & $\beta_{1,1}$ = -2 & -0.004 (0.065) & -0.011 (0.050) & -0.013 (0.107) \\
\multicolumn{1}{l}{State 2} &       &       &       &       & \multicolumn{1}{l}{State 2} &       &       &       &  \\
\multicolumn{1}{l}{j=1} & $\beta_{0,2}$ = 3 & 0.022 (0.102) & 0.011 (0.055) & 0.013 (0.086) & \multicolumn{1}{l}{j=1} & $\beta_{0,2}$ = 3 & 0.009 (0.102) & 0.007 (0.055) & 0.007 (0.084) \\
      & $\beta_{1,2}$ = -2 & 0.006 (0.093) & 0.007 (0.058) & 0.014 (0.087) &       & $\beta_{1,2}$ = -2 & -0.004 (0.093) & 0.005 (0.057) & 0.011 (0.087) \\
\multicolumn{1}{l}{j=2} & $\beta_{0,2}$ = -2 & 0.006 (0.085) & 0.000 (0.054) & 0.003 (0.098) & \multicolumn{1}{l}{j=2} & $\beta_{0,2}$ = -2 & 0.005 (0.085) & 0.001 (0.055) & 0.010 (0.104) \\
      & $\beta_{1,2}$ = 1 & -0.009 (0.087) & 0.000 (0.059) & 0.000 (0.099) &       & $\beta_{1,2}$ = 1 & -0.009 (0.086) & 0.002 (0.058) & 0.010 (0.105) \\
\bottomrule
\end{tabular}%
}%
        \caption{Bias and standard error values of the state-regression parameter estimates for CEHMM with skew t distributed errors for $T = 500$ (Panel A) and $T = 1000$ (Panel B).}
  \label{tab:sim_stdist_e}%
\end{table}%

\section{Empirical Application}\label{sec:emp}
In this section we apply the proposed CQHMM and CEHMM to analyze daily returns of the five major cryptocurrencies as functions of global market indices. The goal of the analysis is to investigate their relationship with leading market indices and describe their dependence structure at different volatility states.
\subsection{Descriptive Statistics}
\begin{figure}[htbp]
\centering
    \begin{subfigure}{1.0\textwidth}
        \includegraphics[width=\linewidth]{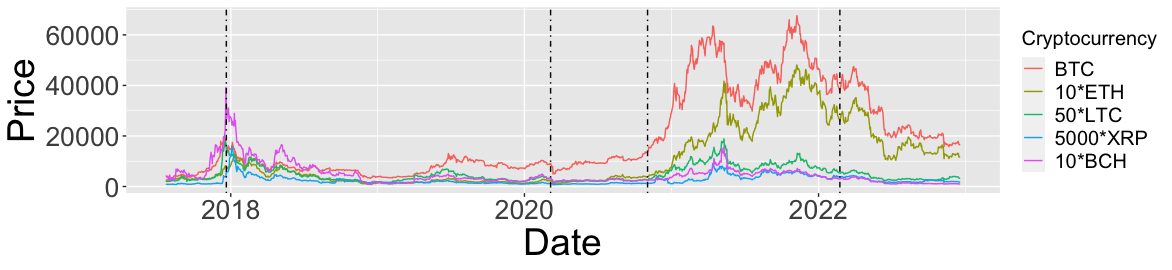}
        \label{fig:prices}
    \end{subfigure}
	 \begin{subfigure}{1.0\textwidth}
        \includegraphics[width=\linewidth]{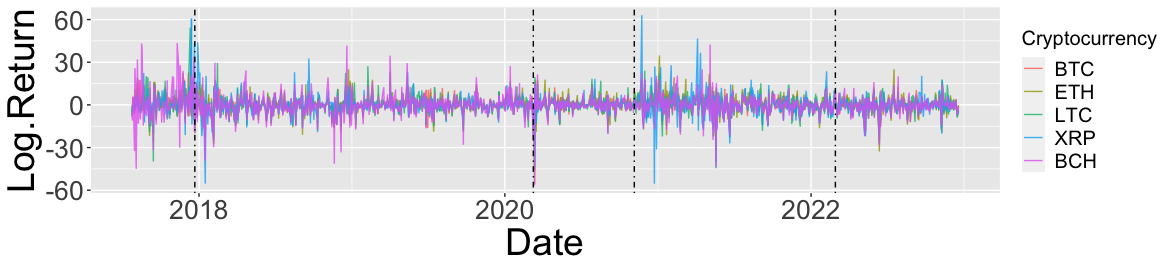}
        \label{fig:returns}
    \end{subfigure}
	\caption{Cryptocurrencies daily prices (top) and log return (bottom) series. Vertical dashed lines indicate globally relevant events in the financial markets that occurred in 2017,12; 2020,03; 2020,11; and 2022,02. Prices are multiplied by a constant to have a similar scale.}
        \label{fig:series}
\end{figure}

 Following \cite{pennoni2022exploring}, we choose crypto-assets that met some requirements of scarcity and tradability on reliable exchanges. In light of these decisions, we consider Bitcoin (BTC), Ethereum (ETH), Litecoin (LTC), Ripple (XRP), Bitcoin cash (BCH) as dependent variables. To investigate for the interlinkages between the crypto markets and non-crypto global markets, we select the S$\&$P500 (GSPC), S$\&$P US Treasury Bond (SPUSBT), US dollar index (USDX), WTI crude oil and Gold as independent variables. The considered timespan extends from July, 25 2017 to December, 19 2022, including numerous crises that have impacted cross-market integration patterns, such as the crypto price bubbles of early 2018, the COVID-19 pandemic, Biden's election at the USA presidency in November 2020 and the Russian invasion of Ukraine at the beginning of 2022, which have caused unprecedented levels of uncertainty and risk. Returns are calculated daily for a total of $T=1348$ observations. The crypto-assets have been downloaded from Coinbase, while the traditional ones from the SPGlobal.com (for S$\&$P500 and S$\&$P US Treasury Bond), Investing.com (for the US dollar index) and Yahoo finance database for the remaining assets.  Figure \ref{fig:series} shows the daily prices (top) and log-returns (bottom) of the five cryptocurrencies over the entire period, where the vertical dotted lines indicate globally relevant events occurred during the study period. We immediately recognise the typical characteristics of this market, i.e. high volatility and sudden waves of exponential price increases.  Daily log-returns of the five cryptocurrencies confirm their high volatility, a strong degree of comovement, and show the typical volatility clustering in common with other traditional financial assets. We observe volatility jumps not only during the first crypto bubble, but also during the financial market crash caused by COVID-19 pandemic and right after Biden's election at the end of 2020, confirming that crypto investors reactions do not differ from the behavior of the investors in the traditional financial markets. 
  In Table \ref{tab:stats} we report the list of examined variables and the summary statistics for the whole sample. The high levels of volatility of cryptocurrencies are noticeable, where Ethereum and Bitcoin Cash in particular stand out, having the highest standard deviation. Crypto assets returns also show very high negative skewness and very high kurtosis, as well as S$\&$P500. The highest level of kurtosis is reported by Crude Oil, which was likely determined by prices fluctuations after the COVID-19 outbreak. On the other hand, the positive skewness of S$\&$P Treasury Bond indicates longer and fatter tails on the right side of the distribution, highlighting an inverse relationship with the S$\&$P500.
In concluding, the Augmented Dickey-Fuller (ADF) test \cite{dickey1979distribution} shows that all daily returns are stationary at the $1\%$ level of significance. 
 The bottom part of Table \ref{tab:stats} also reports the empirical correlations among the dependent variables. As expected, the five cryptocurrencies are all highly correlated, justifying the multivariate approach of the paper.

 Following these considerations, the proposed copula-based QHMM and EHMM are able to provide useful insights on the evolution of the relationships within crypto assets and between non-crypto markets under different market conditions. 

\begin{table}[htbp]
  \centering
  \scalebox{0.7}{
\begin{tabular}{lrrrrrrrr}
      & \multicolumn{1}{l}{Min} & \multicolumn{1}{l}{Mean} & \multicolumn{1}{l}{Max} & \multicolumn{1}{l}{Stdev} & \multicolumn{1}{l}{Skewness} & \multicolumn{1}{l}{Kurtosis} & \multicolumn{1}{l}{Jarque-Bera test} & \multicolumn{1}{l}{ADF test} \\
\midrule
BTC   & -46.47 & 0.13  & 22.51 & 4.89  & -0.77 & 8.79  & \textbf{4471.41} & \textbf{-9.44} \\
ETH   & -55.07 & 0.12  & 34.35 & 6.28  & -0.66 & 7.42  & \textbf{3186.99} & \textbf{-9.57} \\
LTC   & -44.91 & 0.03  & 53.98 & 6.62  & -0.03 & 9.05  & \textbf{4602.24} & \textbf{-9.89} \\
XRP   & -55.05 & 0.04  & 62.67 & 7.60   & 0.74  & 15.10  & \textbf{12927.1} & \textbf{-9.49} \\
BCH   & -56.13 & -0.11 & 43.16 & 8.06  & 0.00  & 8.08  & \textbf{3670.89} & \textbf{-9.89} \\
GSPC  & -12.77 & 0.04  & 8.97  & 1.33  & -0.83 & 14.11 & \textbf{11340.81} & \textbf{-9.87} \\
SPUSBT & -1.69 & 0.00  & 1.79  & 0.28  & 0.15  & 5.10   & \textbf{1467.35} & \textbf{-10.14} \\
USDX  & -2.17 & 0.01  & 2.10   & 0.43  & 0.02  & 2.01  & \textbf{226.45} & \textbf{-11.38} \\
WTI    & -28.22 & 0.08  & 31.96 & 3.29  & 0.04  & 24.70  & \textbf{34272.07} & \textbf{-8.09} \\
GOLD    & -5.11 & 0.03  & 5.78  & 0.94  & -0.20  & 4.95  & \textbf{1387.37} & \textbf{-10.76} \\
      &       &       &       &       &       &       &       &  \\
Correlation matrix \\
 & ETH   & LTC   & XRP   & BCH   &       &       &       &  \\
BTC   & 0.76  & 0.74  & 0.52  & 0.63  &       &       &       &  \\
ETH   &       & 0.80  & 0.62  & 0.68  &       &       &       &  \\
LTC   &       &       & 0.59  & 0.66  &       &       &       &  \\
XRP   &       &       &       & 0.53  &       &       &       &  \\
\bottomrule
\end{tabular}}%
      \caption{Descriptive statistics for the whole sample. The Jarque-Bera test and the ADF test statistics are displayed in boldface when the null hypothesis is rejected at the 1$\%$ significance level.}
  \label{tab:stats}%
\end{table}%

\subsection{Main Results}
In order to apply the aforementioned methods to cryptocurrency returns we consider the model in \eqref{eq:linmod} as:

\be 
\bs \mu_{t,k}^{Crypto} = \bs x_{t-1} \bs \beta_k (\bs \tau),
\ee
where $\bs{\mu}_{tk}^{Crypto} = ({\mu}_{tk}^{BTC}, {\mu}_{tk}^{ETH}, \mu_{tk}^{LTC}, \mu_{tk}^{XRP}, \mu_{tk}^{BCH})$ denotes the vector of all component-wise quantile (expectile) regression functions of the cryptos considered (BTC, ETH, LTC, XRP, BCH) at time $t$ for a given state $k = 1,\dots,K$, $\bs x_{t-1}$ is the vector of returns at the previous date for the S$\&$P500, SPUSBT, USDX, WTI and Gold, with the first element being the intercept, and $\bs \beta_k (\bs \tau)$ is the $p \times d$ matrix of state-dependent regression coefficients defined as $\bs \beta_k (\bs \tau) = (\bs \beta_{1,k} (\tau_1), \dots, \bs \beta_{d,k} (\tau_d))$.

\begin{figure}[htbp]
\centering
\includegraphics[width=0.495\linewidth]{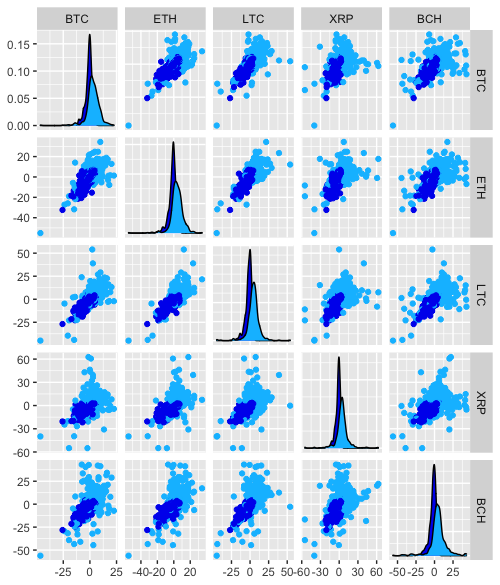}
\includegraphics[width=0.495\linewidth]{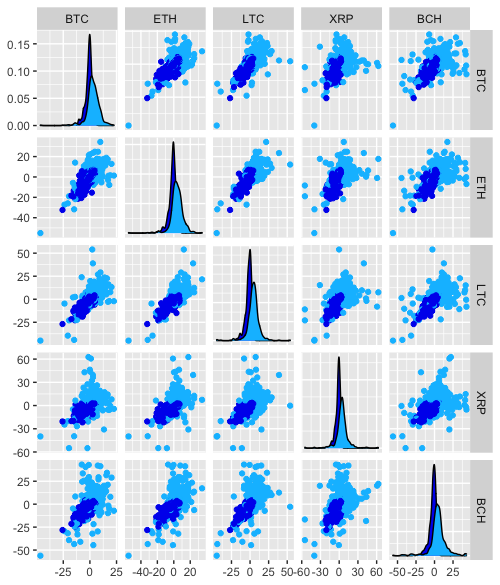}
\caption{Cryptocurrencies marginal distributions and scatterplots classified according to the estimated posterior probability for CQHMM (left) and CEHMM (right) of class membership for $\tau = 0.5$. Dark-blue points (State 1) identify low volatility periods, while light-blue ones (State 2) identify high volatility periods.}
        \label{fig:scatter}
\end{figure}

The first step of the empirical analysis requires the choice of the appropriate copula and number of states. In order to do so, we fit the proposed CQHMM and CEHMM using the Gaussian and t copulas described in Section \ref{sec:method} for a grid of different values of $K$, spanning from 1 to 4. For ease of comparison between the two models, the copula function and $K$ are selected for $\tau = \tau_j = 0.50$, $j=1,\dots,d$. Then, in the analysis we have kept this choice fixed when fitting the models at the other values of $\tau_j \neq 0.50$, $j=1,\dots,d$.  We consider three widely used penalized likelihood selection criteria, namely the AIC \citep{akaike1998information}, the BIC \citep{schwarz1978estimating} and the ICL \citep{biernacki2000assessing}, and report the results in Table \ref{tab:criteria}. 
 In order to clearly identify high and low volatility market conditions, we use $K=2$ which is supported by the parsimonious ICL criteria for both CQHMM and CEHMM, together with a t copula. From a graphical perspective, in Figure \ref{fig:scatter} we report the scatterplots and the marginal densities colored according to the estimated posterior probability of class membership, $\underset{k}{\max} \, \gamma_t (k)$.

 We thus fit the CQHMM and CEHMM under the t copula for $K=2$ hidden states at three levels $\tau = \{0.05, 0.5, 0.95\}$, with $\tau = \tau_j$ for all $j=1,\dots,d$. The estimates of the state-specific parameters are gathered in Tables \ref{tab:betas_tau05_q}, \ref{tab:betas_tau5_q} and \ref{tab:betas_tau95_q} for the CQHMM and in Tables \ref{tab:betas_tau05_e}, \ref{tab:betas_tau5_e} and  \ref{tab:betas_tau95_e} for the CEHMM, along with the standard errors (in brackets), computed by using the parametric bootstrap technique illustrated in Section \ref{sec:method} over $R = 1000$ resamples.
 As regards the estimated scale parameters, $\sigma_1$ reflects stable periods, representing the so-called low-volatility state, meanwhile $\sigma_2$ contemplates rapid (positive and negative) peak and burst returns, which defines the second state as the high-volatility state. These results confirm the graphical analysis conducted in Figure \ref{fig:scatter}. By looking at the state-dependent degrees of freedom in each table mentioned above, it is clear the necessity to forgo the Gaussian copula in favor of the more robust, fat-tailed t copula for dependence modeling in financial data.
 Taking the impact of covariates into account, we first comment on the parameter estimates of the CQHMM (see Tables \ref{tab:betas_tau05_q}, \ref{tab:betas_tau5_q} and \ref{tab:betas_tau95_q}). As could be expected, the state-specific intercepts are increasing somewhat with $\tau$, with State 1 having lower absolute values than State 2 for all $\tau$'s. For $\tau = 0.50$ (see Table \ref{tab:betas_tau5_q}), we observe that at low volatility periods S$\&$P500 is the only statistically significant asset, negatively influencing almost all the cryptocurrencies for both quantile and expectile models, which implies that during tranquil periods crypto assets can be considered as weak hedges \citep{bouri2017bitcoin, bouri2017hedge}.
 During high volatility periods we observe that S$\&$P500, Gold and Crude Oil influence some cryptos, especially Bitcoin and Litecoin, while in the CEHMM (Table \ref{tab:betas_tau5_e}) there is no statistical association among cryptocurrencies and the assets considered. 
 The number of significantly parameters increases by moving to the extreme tails of cryptocurrencies returns distribution, exposing a connection during bearish and bullish market periods between traditional financial markets and the crypto market. In particular, for the CQHMM at $\tau = 0.05$ (Table \ref{tab:betas_tau05_q}) we observe that at low volatility periods Gold negatively influences all the cryptos considered. It can also be seen that the negative impact of S$\&$P Treasury Bond is statistically significant with respect to all crypto-assets considered, with the exception of Ethereum. In the second state, most of the regression parameters are significantly different from zero, with few exceptions. We highlight in particular the strong positive influence of S$\&$P Treasury Bond and the negative one of US dollar index. The CEHMM shows similar results at $\tau = 0.05$ for the second state (Table \ref{tab:betas_tau05_e}), where we highlight the strong positive influences of S$\&$P500 and S$\&$P Treasury Bond, with the exception of Ripple and Bitcoin Cash. Finally, at $\tau = 0.95$ for the CQHMM (Table \ref{tab:betas_tau95_q}) for both states considered we note strong influences of S$\&$P Treasury Bond, US dollar index and Gold. Similar results can be highlighted for the CEHMM (Table \ref{tab:betas_tau95_e}) in the high volatility state, especially regarding the strong negative influences of US dollar index and Gold.
 Overall, these results are consistent with the recent works of \cite{bouri2020cryptocurrencies, conlon2020safe, corbet2020contagion, caferra2021raised} and \cite{yousaf2021linkages} and highlight that: (i) the relationship among these two different markets is rather complex, and it is more pronounced in the tails of the returns distributions; (ii) the dependence within the crypto market varies over time according to the market conditions.
 
\begin{figure}
	\centering
	\includegraphics[width=.449\linewidth]{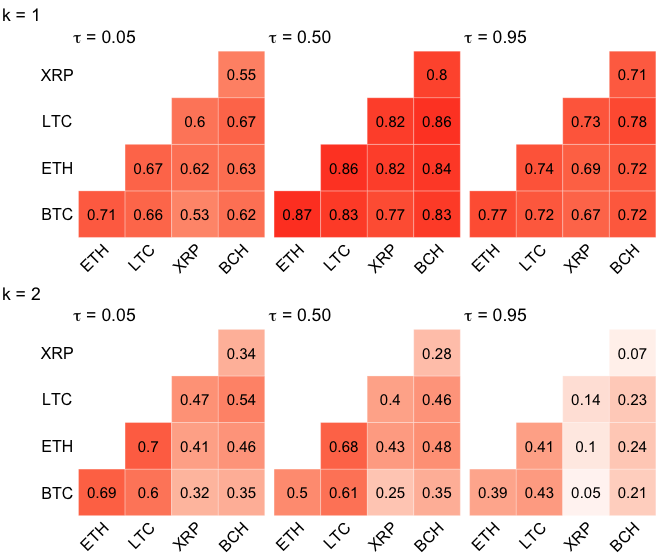}
	\includegraphics[width=.495\linewidth]{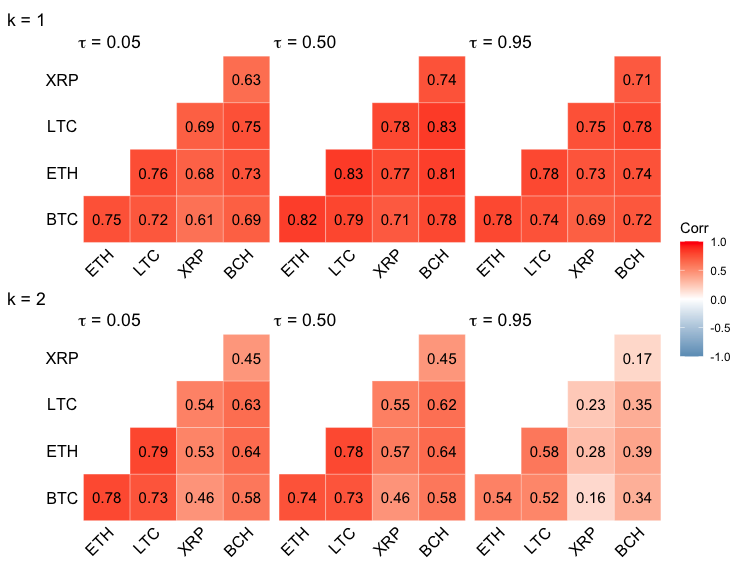}
	\caption{CQHMM (left) and CEHMM (right) state-specific correlation estimates for the two states at $\tau = \{0.05, 0.50, 0.95\}$.}
        \label{fig:corrplots}
\end{figure}

 Finally, Figure \ref{fig:corrplots} reports the estimated pair-wise correlations of the t copula for both states under the CQHMM and CEHMM models at $\tau = \{0.05, 0.50, 0.95\}$.  Overall we observe higher correlation estimates during low volatility states, especially at $\tau = 0.50$. Looking at specific values, we highlight the high correlation levels between BTC, ETH and LTC and, on the other hand it is visible the divergent behavior of XRP with respect to the other cryptocurrencies \citep{pennoni2022exploring}. 
  
\begin{table}[htbp]
  \centering
  \scalebox{0.8}{
\begin{tabular}{lrrrrrr}
      & \multicolumn{3}{c}{Gaussian copula} & \multicolumn{3}{c}{t copula} \\ \cmidrule(l){2-4} \cmidrule(l){5-7}
      & AIC   & BIC   & ICL   & AIC   & BIC   & ICL \\
\midrule
\textbf{CQHMM} &       &       &       &       &       &  \\
$K=1$   & 37322.29 & 37556.55 & 37556.55 & 35970.66 & 36210.11 & 36210.11 \\
$K=2$   & 35326.37 & \textbf{35810.50} & \textbf{36137.40} & 35160.71 & \textbf{35655.25} & \textbf{35821.08} \\
$K=3$   & 35078.00 & 35822.41 & 36609.45 & 34988.46 & 35748.48 & 36655.90 \\
$K=4$   & \textbf{34871.65} & 35886.75 & 36908.26 & \textbf{34802.28} & 35838.20 & 36657.36 \\
\\
\textbf{CEHMM} &       &       &       &       &       &  \\
$K=1$   & 35157.65 & 35362.76 & 35362.76 & 34455.43 & 34665.67 & 34665.67 \\
$K=2$   & 32354.36 & 32779.96 & \textbf{33075.53} & 32308.68 & 32744.53 & \textbf{33067.44} \\
$K=3$   & 31869.51 & \textbf{32525.86} & 33140.76 & 31871.03 & \textbf{32542.76} & 33203.10 \\
$K=4$   & \textbf{31635.38} & 32532.72 & 33390.67 & \textbf{31642.18} & 32560.04 & 33387.36 \\
\bottomrule
\end{tabular}%

}%
    \caption{AIC, BIC and ICL values with varying number of states for the CQHMM and CEHMM under the Gaussian and t copulas. Bold font highlights the best values for the considered criteria (lower-is-better).}
    \label{tab:criteria}%
\end{table}%

\begin{table}[htbp]
  \centering
  \scalebox{0.55}{
\begin{tabular}{lrrrrr}
\textbf{CQHMM} & BTC   & ETH   & LTC   & XRP   & BCH \\
\midrule
State 1 &       &       &       &       &  \\
Intercept & \textbf{-2.65 (0.038)} & \textbf{-3.575 (0.051)} & \textbf{-3.957 (0.055)} & \textbf{-3.926 (0.051)} & \textbf{-4.205 (0.056)} \\
GSPC  & \textbf{-0.166 (0.035)} & -0.075 (0.047) & \textbf{-0.149 (0.051)} & -0.095 (0.051) & \textbf{-0.148 (0.053)} \\
SPUSBT & \textbf{-0.716 (0.166)} & -0.061 (0.218) & \textbf{-1.897 (0.234)} & \textbf{-1.078 (0.240)} & \textbf{-1.117 (0.244)} \\
USDX  & 0.109 (0.108) & -0.261 (0.150) & \textbf{-0.609 (0.159)} & \textbf{-0.732 (0.154)} & \textbf{-0.714 (0.167)} \\
WTI   & -0.008 (0.015) & \textbf{-0.079 (0.020)} & \textbf{-0.084 (0.021)} & \textbf{-0.148 (0.02)} & \textbf{-0.089 (0.023)} \\
GOLD  & \textbf{-0.138 (0.051)} & \textbf{-0.428 (0.066)} & \textbf{-0.175 (0.072)} & \textbf{-0.377 (0.071)} & \textbf{-0.44 (0.077)} \\
$\sigma_1$ & 0.237 (0.008) & 0.313 (0.011) & 0.34 (0.011) & 0.328 (0.011) & 0.353 (0.012) \\
$\nu_1$ & 5.022 (0.505) &       &       &       &  \\
      &       &       &       &       &  \\
State 2 &       &       &       &       &  \\
Intercept & \textbf{-14.552 (0.224)} & \textbf{-19.013 (0.297)} & \textbf{-18.956 (0.294)} & \textbf{-19.268 (0.361)} & \textbf{-22.306 (0.401)} \\
GSPC  & \textbf{0.809 (0.230)} & 0.567 (0.292) & \textbf{1.765 (0.296)} & \textbf{1.172 (0.364)} & \textbf{2.639 (0.430)} \\
SPUSBT & \textbf{7.348 (1.031)} & \textbf{7.462 (1.429)} & \textbf{3.938 (1.356)} & 1.588 (1.704) & -1.034 (1.934) \\
USDX  & \textbf{1.402 (0.685)} & \textbf{-3.617 (0.866)} & -1.315 (0.895) & \textbf{-4.287 (1.108)} & \textbf{-2.4 (1.202)} \\
WTI   & \textbf{0.472 (0.097)} & \textbf{0.908 (0.129)} & \textbf{0.403 (0.129)} & \textbf{0.757 (0.155)} & -0.132 (0.181) \\
GOLD  & 0.088 (0.326) & \textbf{-0.872 (0.425)} & \textbf{1.223 (0.433)} & -0.546 (0.552) & 0.414 (0.589) \\
$\sigma_2$ & 0.692 (0.040) & 0.909 (0.056) & 0.898 (0.054) & 1.102 (0.064) & 1.24 (0.073) \\
$\nu_2$ & 7.867 (2.485) &       &       &       &  \\
\bottomrule
\end{tabular}}%
    \caption{CQHMM state-specific parameter estimates for $\tau = 0.05$, with bootstrapped standard errors (in brackets) obtained over 1000 replications. Point estimates are displayed in boldface when significant at the standard 5\% level. $\sigma_k$ and $\nu_k$ represent the state-specific scale parameter and degrees of freedom, respectively. }
  \label{tab:betas_tau05_q}%
\end{table}%

\begin{table}[htbp]
  \centering
  \scalebox{0.55}{
\begin{tabular}{lrrrrr}
\textbf{CQHMM} & BTC   & ETH   & LTC   & XRP   & BCH \\
\midrule
State 1 &       &       &       &       &  \\
Intercept & 0.029 (0.096) & -0.125 (0.125) & -0.187 (0.125) & \textbf{-0.249 (0.117)} & \textbf{-0.383 (0.132)} \\
GSPC  & \textbf{-0.197 (0.083)} & \textbf{-0.307 (0.108)} & -0.143 (0.114) & \textbf{-0.227 (0.098)} & \textbf{-0.374 (0.110)} \\
SPUSBT & -0.011 (0.405) & -0.54 (0.546) & -0.341 (0.524) & -0.123 (0.511) & -0.275 (0.538) \\
USDX  & 0.093 (0.264) & 0.038 (0.342) & 0.036 (0.344) & 0.151 (0.332) & -0.103 (0.349) \\
WTI   & -0.005 (0.034) & -0.029 (0.044) & -0.07 (0.047) & 0.006 (0.041) & 0.003 (0.044) \\
GOLD  & -0.057 (0.121) & -0.09 (0.158) & -0.175 (0.158) & -0.072 (0.152) & -0.072 (0.162) \\
$\sigma_1$ & 1.485 (0.046) & 1.975 (0.061) & 2 (0.061) & 1.842 (0.056) & 2.014 (0.060) \\
$\nu_1$ & 5.73 (0.625) &       &       &       &  \\
      &       &       &       &       &  \\
State 2 &       &       &       &       &  \\
Intercept & \textbf{1.438 (0.314)} & \textbf{1.198 (0.399)} & \textbf{1.556 (0.457)} & 1.209 (0.651) & \textbf{2.263 (0.723)} \\
GSPC  & \textbf{0.58 (0.292)} & \textbf{0.922 (0.376)} & \textbf{1.077 (0.413)} & 0.099 (0.605) & -0.122 (0.658) \\
SPUSBT & 0.192 (1.447) & 1.311 (1.731) & -0.286 (1.880) & -1.741 (2.822) & 0.738 (3.045) \\
USDX  & 0.163 (0.945) & 1.04 (1.130) & 1.823 (1.271) & -3.139 (1.876) & 0.566 (2.027) \\
WTI   & \textbf{0.262 (0.133)} & 0.148 (0.152) & \textbf{0.497 (0.178)} & 0.232 (0.256) & \textbf{0.649 (0.283)} \\
GOLD  & -0.364 (0.429) & -0.06 (0.542) & \textbf{1.196 (0.601)} & -1.509 (0.812) & -0.901 (0.934) \\
$\sigma_2$ & 2.231 (0.145) & 2.739 (0.174) & 3.06 (0.190) & 4.456 (0.284) & 4.911 (0.312) \\
$\nu_2$ & 8.276 (2.823) &       &       &       &  \\
\bottomrule
\end{tabular}}%
    \caption{CQHMM state-specific parameter estimates for $\tau = 0.50$, with bootstrapped standard errors (in brackets) obtained over 1000 replications. Point estimates are displayed in boldface when significant at the standard 5\% level. $\sigma_k$ and $\nu_k$ represent the state-specific scale parameter and degrees of freedom, respectively. }
  \label{tab:betas_tau5_q}%
\end{table}%

\begin{table}[htbp]
  \centering
  \scalebox{0.55}{
\begin{tabular}{lrrrrr}
\textbf{CQHMM} & BTC   & ETH   & LTC   & XRP   & BCH \\
\midrule
State 1 &      &       &       &       &  \\
Intercept & \textbf{3.062 (0.051)} & \textbf{4.19 (0.064)} & \textbf{3.008 (0.057)} & \textbf{2.703 (0.054)} & \textbf{2.805 (0.058)} \\
GSPC  & \textbf{0.103 (0.047)} & \textbf{0.139 (0.063)} & \textbf{-0.313 (0.053)} & \textbf{-0.321 (0.050)} & \textbf{-0.342 (0.055)} \\
SPUSBT & \textbf{1.607 (0.219)} & \textbf{2.133 (0.287)} & -0.152 (0.255) & -0.107 (0.243) & 0.387 (0.273) \\
USDX  & \textbf{0.396 (0.145)} & \textbf{0.584 (0.180)} & \textbf{-0.747 (0.163)} & \textbf{-0.409 (0.146)} & -0.115 (0.170) \\
WTI   & \textbf{0.056 (0.020)} & -0.033 (0.025) & 0.004 (0.022) & -0.001 (0.021) & -0.013 (0.023) \\
GOLD  & \textbf{-0.378 (0.065)} & \textbf{-0.557 (0.082)} & 0.047 (0.076) & -0.112 (0.070) & -0.125 (0.081) \\
$\sigma_1$ & 0.268 (0.009) & 0.354 (0.013) & 0.31 (0.011) & 0.294 (0.010) & 0.324 (0.011) \\
$\nu_1$ & 3.382 (0.311) &       &       &       &  \\
      &       &       &       &       &  \\
State 2 &       &       &       &       &  \\
Intercept & \textbf{10.69 (0.142)} & \textbf{13.979 (0.182)} & \textbf{14.249 (0.203)} & \textbf{17.824 (0.305)} & \textbf{21.306 (0.300)} \\
GSPC  & -0.034 (0.138) & 0.182 (0.176) & 0.034 (0.203) & \textbf{-2.569 (0.313)} & -0.236 (0.289) \\
SPUSBT & -0.064 (0.672) & 1.051 (0.807) & \textbf{1.908 (0.895)} & \textbf{-11.908 (1.414)} & \textbf{-9.343 (1.356)} \\
USDX  & -0.79 (0.405) & \textbf{3.829 (0.538)} & \textbf{-1.667 (0.615)} & \textbf{-4.206 (0.906)} & \textbf{-4.33 (0.896)} \\
WTI   & \textbf{-0.181 (0.058)} & 0.116 (0.071) & 0.017 (0.078) & -0.028 (0.131) & 0.184 (0.124) \\
GOLD  & \textbf{-0.697 (0.196)} & -0.458 (0.242) & \textbf{-2.435 (0.287)} & \textbf{1.839 (0.430)} & \textbf{-2.912 (0.422)} \\
$\sigma_2$ & 0.565 (0.026) & 0.707 (0.033) & 0.808 (0.038) & 1.228 (0.055) & 1.195 (0.053) \\
$\nu_2$ & 6.265 (0.956) &       &       &       &  \\
\bottomrule
\end{tabular}}%
    \caption{CQHMM state-specific parameter estimates for $\tau = 0.95$, with bootstrapped standard errors (in brackets) obtained over 1000 replications. Point estimates are displayed in boldface when significant at the standard 5\% level. $\sigma_k$ and $\nu_k$ represent the state-specific scale parameter and degrees of freedom, respectively. }
  \label{tab:betas_tau95_q}%
\end{table}%

\begin{table}[htbp]
  \centering
  \scalebox{0.55}{
\begin{tabular}{lrrrrr}
\textbf{CEHMM} & BTC   & ETH   & LTC   & XRP   & BCH \\
\midrule
State 1 &       &       &       &       &  \\
Intercept & \textbf{-2.666 (0.085)} & \textbf{-3.448 (0.106)} & \textbf{-3.755 (0.108)} & \textbf{-3.674 (0.106)} & \textbf{-3.894 (0.109)} \\
GSPC  & \textbf{-0.164 (0.072)} & -0.134 (0.093) & -0.051 (0.094) & -0.097 (0.092) & -0.14 (0.092) \\
SPUSBT & \textbf{-0.675 (0.340)} & -0.79 (0.425) & -0.567 (0.412) & -0.6 (0.416) & -0.132 (0.434) \\
USDX  & 0.344 (0.229) & -0.254 (0.280) & 0.124 (0.291) & -0.384 (0.278) & -0.336 (0.299) \\
WTI   & -0.019 (0.029) & -0.041 (0.036) & -0.05 (0.037) & \textbf{-0.076 (0.035)} & -0.059 (0.037) \\
GOLD  & \textbf{0.223 (0.104)} & 0.181 (0.130) & 0.173 (0.131) & 0.065 (0.127) & 0.157 (0.137) \\
$\sigma_1$ & 1.812 (0.041) & 2.247 (0.051) & 2.33 (0.053) & 2.225 (0.052) & 2.322 (0.052) \\
$\nu_1$ & 6.985 (0.833) &       &       &       &  \\
      &       &       &       &       &  \\
State 2 &       &       &       &       &  \\
Intercept & \textbf{-12.027 (0.495)} & \textbf{-15.603 (0.626)} & \textbf{-15.652 (0.670)} & \textbf{-16.562 (0.848)} & \textbf{-19.073 (0.844)} \\
GSPC  & \textbf{1.743 (0.416)} & \textbf{1.719 (0.546)} & \textbf{2.097 (0.594)} & 1.470 (0.775) & \textbf{3.203 (0.769)} \\
SPUSBT & \textbf{8.728 (1.938)} & \textbf{10.294 (2.685)} & \textbf{6.881 (2.717)} & 2.749 (3.616) & 2.861 (3.713) \\
USDX  & -0.612 (1.325) & -1.180 (1.696) & 0.402 (1.864) & 0.016 (2.361) & 3.631 (2.448) \\
WTI   & \textbf{0.684 (0.177)} & \textbf{0.879 (0.227)} & \textbf{0.621 (0.251)} & \textbf{0.818 (0.317)} & 0.116 (0.319) \\
GOLD  & 0.641 (0.595) & 0.755 (0.798) & 1.652 (0.847) & 1.057 (1.095) & 1.775 (1.132) \\
$\sigma_2$ & 4.66 (0.217) & 6.13 (0.294) & 6.461 (0.302) & 8.436 (0.388) & 8.442 (0.395) \\
$\nu_2$ & 14.816 (93.272) &       &       &       &  \\
\bottomrule
\end{tabular}}%
    \caption{CEHMM state-specific parameter estimates for $\tau = 0.05$, with bootstrapped standard errors (in brackets) obtained over 1000 replications. Point estimates are displayed in boldface when significant at the standard 5\% level. $\sigma_k$ and $\nu_k$ represent the state-specific scale parameter and degrees of freedom, respectively. }
  \label{tab:betas_tau05_e}%
\end{table}%

\begin{table}[htbp]
  \centering
  \scalebox{0.55}{
\begin{tabular}{lrrrrr}
\textbf{CEHMM} & BTC   & ETH   & LTC   & XRP   & BCH \\
\midrule
State 1 &       &       &       &       &  \\
Intercept & 0.054 (0.109) & -0.019 (0.137) & \textbf{-0.358 (0.132)} & \textbf{-0.402 (0.127)} & \textbf{-0.547 (0.136)} \\
GSPC  & \textbf{-0.228 (0.091)} & \textbf{-0.243 (0.117)} & \textbf{-0.232 (0.111)} & -0.199 (0.106) & \textbf{-0.261 (0.117)} \\
SPUSBT & -0.134 (0.454) & -0.173 (0.576) & -0.677 (0.545) & -0.339 (0.525) & 0.093 (0.563) \\
USDX  & -0.007 (0.284) & -0.123 (0.369) & -0.113 (0.351) & -0.061 (0.338) & -0.160 (0.355) \\
WTI   & 0.029 (0.036) & 0.008 (0.047) & 0.001 (0.044) & -0.023 (0.043) & 0.008 (0.045) \\
GOLD  & -0.051 (0.131) & -0.101 (0.166) & -0.052 (0.161) & 0.005 (0.154) & -0.162 (0.162) \\
$\sigma_1$ & 3.388 (0.081) & 4.351 (0.101) & 4.136 (0.097) & 3.981 (0.095) & 4.227 (0.098) \\
$\nu_1$ & 11.101 (2.090) &       &       &       &  \\
      &       &       &       &       &  \\
State 2 &       &       &       &       &  \\
Intercept & 0.467 (0.439) & 0.678 (0.570) & \textbf{1.339 (0.613)} & \textbf{1.46 (0.731)} & 1.400 (0.790) \\
GSPC  & 0.294 (0.381) & 0.31 (0.497) & 0.326 (0.547) & -0.137 (0.656) & 0.610 (0.689) \\
SPUSBT & 1.355 (1.816) & 3.534 (2.264) & 1.874 (2.603) & -1.448 (3.076) & -2.041 (3.310) \\
USDX  & 0.767 (1.182) & 1.861 (1.550) & 1.493 (1.748) & -1.250 (2.062) & 0.698 (2.120) \\
WTI   & -0.016 (0.153) & 0.043 (0.195) & 0.156 (0.219) & 0.129 (0.262) & 0.106 (0.282) \\
GOLD  & -0.233 (0.560) & -0.121 (0.714) & -0.37 (0.775) & -0.305 (0.911) & -0.810 (0.995) \\
$\sigma_2$ & 7.822 (0.318) & 10.026 (0.410) & 11.074 (0.459) & 13.522 (0.527) & 14.285 (0.576) \\
$\nu_2$ & 12.806 (9.207) &       &       &       &  \\
\bottomrule
\end{tabular}}%
    \caption{CEHMM state-specific parameter estimates for $\tau = 0.50$, with bootstrapped standard errors (in brackets) obtained over 1000 replications. Point estimates are displayed in boldface when significant at the standard 5\% level. $\sigma_k$ and $\nu_k$ represent the state-specific scale parameter and degrees of freedom, respectively. }
  \label{tab:betas_tau5_e}%
\end{table}%

\begin{table}[htbp]
  \centering
  \scalebox{0.55}{
\begin{tabular}{lrrrrr}
\textbf{CEHMM} & BTC   & ETH   & LTC   & XRP   & BCH \\
\midrule
State 1 &       &       &       &       &  \\
Intercept & \textbf{2.466 (0.090)} & \textbf{3.443 (0.121)} & \textbf{2.499 (0.108)} & \textbf{2.132 (0.102)} & \textbf{2.388 (0.114)} \\
GSPC  & \textbf{-0.153 (0.073)} & -0.177 (0.099) & \textbf{-0.296 (0.091)} & \textbf{-0.327 (0.084)} & \textbf{-0.247 (0.096)} \\
SPUSBT & \textbf{0.924 (0.359)} & \textbf{1.275 (0.483)} & -0.087 (0.434) & 0.008 (0.394) & 0.298 (0.447) \\
USDX  & -0.056 (0.229) & -0.572 (0.318) & -0.378 (0.282) & -0.432 (0.259) & -0.532 (0.291) \\
WTI   & 0.035 (0.029) & -0.029 (0.039) & 0.016 (0.035) & -0.001 (0.033) & -0.014 (0.036) \\
GOLD  & -0.124 (0.109) & -0.205 (0.145) & -0.037 (0.132) & -0.093 (0.122) & -0.248 (0.137) \\
$\sigma_1$ & 1.731 (0.041) & 2.383 (0.055) & 2.152 (0.049) & 1.981 (0.045) & 2.234 (0.052) \\
$\nu_1$ & 5.753 (0.673) &       &       &       &  \\
      &       &       &       &       &  \\
State 2 &       &       &       &       &  \\
Intercept & \textbf{9.344 (0.291)} & \textbf{11.905 (0.345)} & \textbf{13.356 (0.448)} & \textbf{17.590 (0.621)} & \textbf{17.766 (0.593)} \\
GSPC  & 0.023 (0.253) & 0.236 (0.329) & 0.401 (0.385) & -1.030 (0.608) & \textbf{-0.991 (0.490)} \\
SPUSBT & -0.692 (1.199) & \textbf{4.962 (1.553)} & 3.312 (1.865) & -3.591 (2.733) & -2.229 (2.527) \\
USDX  & -0.206 (0.807) & \textbf{3.267 (0.999)} & -0.244 (1.264) & \textbf{-6.721 (1.789)} & -2.002 (1.674) \\
WTI   & -0.137 (0.109) & 0.034 (0.127) & 0.105 (0.160) & 0.074 (0.228) & 0.322 (0.197) \\
GOLD  & -0.343 (0.363) & -0.707 (0.470) & \textbf{-2.874 (0.573)} & 0.831 (0.845) & \textbf{-1.957 (0.751)} \\
$\sigma_2$ & 3.636 (0.137) & 4.600 (0.173) & 5.645 (0.206) & 8.079 (0.293) & 7.335 (0.269) \\
$\nu_2$ & 6.102 (1.100) &       &       &       &  \\
\bottomrule
\end{tabular}}%
    \caption{CEHMM state-specific parameter estimates for $\tau = 0.95$, with bootstrapped standard errors (in brackets) obtained over 1000 replications. Point estimates are displayed in boldface when significant at the standard 5\% level. $\sigma_k$ and $\nu_k$ represent the state-specific scale parameter and degrees of freedom, respectively. }
  \label{tab:betas_tau95_e}%
\end{table}%

\FloatBarrier

\section{Conclusions}\label{sec:conclusions}
Since cryptocurrencies were first introduced in 2008 as a form of payment and later turned into high yield assets, researchers have questioned their extreme volatility and fluctuations over time. The early 2018 crypto-bubble and the COVID-19 market crash in 2020 have contributed to strengthening the focus on the relationship between digital currencies and traditional financial assets. In this context, our contribution to the existing literature is twofold. From a theoretical standpoint, we grasp unobserved serial heterogeneity and rapid volatility jumps by developing hidden Markov regression models for joint estimation of conditional quantiles or expectiles of multiple time series. At the same time, we consider state-dependent elliptical copula functions to capture the time-varying dependence structure of cryptocurrency returns. 
 
 From a practical point of view, we jointly investigate the impact of global market indices on daily returns of the five most important cryptocurrencies from 2017 to 2022, while taking into account for their association under different market conditions. We found that interrelations between crypto and stocks increase while moving to the extreme tails of returns distributions. In particular, a weak relationship between cryptocurrencies and stock markets and commodities occurr at the centre of the returns' distributions, concurring with the consensus that cryptocurrencies are good diversifiers from stocks and commodities during periods of tranquillity of financial markets \citep{bouri2017hedge, bouri2017bitcoin, cremaschini2022stylized}. Conversely, we highlight an important influence of S$\&$P500, S$\&$P Treasury Bond and Gold during both bearish and bullish periods. Overall, these results are consistent with the existing strands of literature on the subject \citep{bouri2020cryptocurrencies, caferra2021raised, yousaf2021linkages}. Finally, as regards the dependence analysis among the crypto, our results seem to adhere to the ones in \cite{pennoni2022exploring}.
 
Future topics of research could extend the proposed methods to the hidden semi-Markov model setting where the sojourn-distributions, that is, the distributions of the number of consecutive time points that the chain spends in each state, are modeled by the researcher using either parametric or non-parametric approaches instead of assuming geometric sojourn densities as in HMMs. Moreover, even tough the application focused on five cryptocurrencies and five predictors, in high-dimensional settings with a greater number of response variables and/or hidden states, the described models can be easily over-parameterized. This is frequently the case due to the large number of regression parameters and unique parameters in the correlation matrices of the copulas to be estimated, meaning a loss in interpretability as well as numerically ill-conditioned estimators. In these situations, one could specify a parsimonious parametrization of correlation matrices of copula functions or, following \cite{maruotti2017dynamic}, consider mixtures of factor analysis models, whose parameters evolve according to the latent Markov chain.

\newpage
\section*{Acknowledgements}
This research did not receive any specific grant from funding agencies in the public, commercial, or not-for-profit sectors.
\bibliographystyle{agsm}
\bibliography{sn-article.bib}

\end{document}